\documentclass[12pt, draftclsnofoot, onecolumn]{IEEEtran}

\usepackage{bm}
\usepackage{url}
\usepackage{cite}
\usepackage{color}
\usepackage{times}
\usepackage{array}
\usepackage{xcolor}
\usepackage{balance}
\usepackage{amssymb}
\usepackage{amsmath}
\usepackage{comment}
\usepackage{multicol}
\usepackage{graphicx}
\usepackage{marvosym}
\usepackage{multirow}
\usepackage{graphicx}
\usepackage{stfloats}
\usepackage{setspace}
\usepackage{subfigure}
\usepackage{dcolumn,booktabs}
\usepackage{amsmath,amsthm,amssymb,amsfonts}
\usepackage{rotating,threeparttable,booktabs}
\usepackage[ruled,linesnumbered]{algorithm2e}

\newtheorem{remark}{Remark}

\newtheorem{theorem}{Theorem}

\newtheorem{corollary}{Corollary}
\newtheorem{definition}{Definition}
\newtheorem{proposition}{Proposition}

\begin{document}
\begin{spacing}{1.3}

\title{
	{\LARGE Beamforming Design and Performance Evaluation for Reconfigurable Intelligent Surface Assisted Wireless Communication Systems With Non-Ideal Hardware}
}

\author{
	Yiming~Liu,~\IEEEmembership{Student~Member,~IEEE,}
	Erwu~Liu,~\IEEEmembership{Senior~Member,~IEEE,}
    Rui~Wang,~\IEEEmembership{Senior~Member,~IEEE,}
    Binyu~Lu,
    and Zhu~Han,~\IEEEmembership{Fellow,~IEEE}
    
    \thanks{Part of this work is presented in IEEE GLOBECOM 2020\textcolor{blue}{\cite{Liu2012:Energy}}, and part is presented in IEEE WCNC 2021\textcolor{blue}{\cite{Liu2103:Channel}}.
    }

	\thanks{Yiming Liu, Erwu Liu, Rui Wang, and Binyu Lu are with the College of Electronic and Information Engineering, Tongji University, Shanghai 201804, China, E-mail: ymliu\_970131@tongji.edu.cn, erwu.liu@ieee.org.
	}
	
	\thanks{
		Zhu Han is with the Department of Electrical and Computer Engineering, University of Houston, Houston, TX 77004 USA, E-mail: Zhan2@uh.edu.
	}
}

\maketitle

\begin{abstract}
Reconfigurable intelligent surface (RIS) can effectively control the wavefront of the impinging signals and has emerged as a cost-effective promising solution to improve the spectrum and energy efficiency of wireless systems. Most existing researches on RIS assume that the hardware operations are perfect. However, both physical transceiver and RIS suffer from inevitable hardware impairments in practice, which can lead to severe system performance degradation and increase the complexity of beamforming optimization. Consequently, the existing researches on RIS, including channel estimation, beamforming optimization, spectrum and energy efficiency analysis, etc., cannot directly apply to the case of hardware impairments. In this paper, by taking hardware impairments into consideration, we conduct the joint transmit and reflect beamforming optimization, and reevaluate the system performance. First, we characterize the closed-form estimators of direct and cascaded channels in both single-user and multi-user cases, and analyze the impact of hardware impairments on channel estimation accuracy. Then, the optimal transmit beamforming solution is derived, and a gradient descent method-based algorithm is also proposed to optimize the reflect beamforming. Moreover, we analyze the three types of asymptotic channel capacities with respect to the transmit power, the antenna number, and the reflecting element number. Finally, in terms of the system energy consumption, we analyze the power scaling law and the energy efficiency. Our experimental results also reveal an encouraging phenomenon that the RIS-assisted wireless system with massive reflecting elements can achieve both high spectrum and energy efficiency without the need for massive antennas and without allocating too many resources to optimize the reflect beamforming. 
\end{abstract}


\IEEEpeerreviewmaketitle

\section{Introduction}
\IEEEPARstart{T}{he} March 2020 report released by Cisco Systems, Inc., shows that the number of networked devices and connections will reach up to 29.3 billions by the year of 2023, and about half of them are mobile-ready devices and connections\textcolor{blue}{\cite{index2019global}}. 
This inevitably leads to an explosive growth of mobile data traffic, which requires to enhance the performance of wireless communication systems in future. 
The fifth-generation (5G) wireless network technology has been standardized to solve this problem. However, there exists no single enabling technology that can support all 5G application requirements during the standardization process. 
Therefore, it necessitates radically new communication paradigms, especially at the physical layer. Reconfigurable intelligent surface (RIS), {a.k.a.}, intelligent reflecting surface, has emerged as a cost-effective promising method for future wireless communications. Existing works have revealed that, with the assistance of a smart controller, RIS can effectively control the wavefront, {e.g.}, the phase, amplitude, frequency, and even polarization, of the impinging signals without the need of complex decoding, encoding, and radio frequency processing operations, such that the desired signals and interfering signals are added constructively and destructively at the receivers, respectively\textcolor{blue}{\cite{8796365,8888223,8936989,8741198}}. 
Furthermore, RIS is much cheaper than active base station (BS) or relay, and can be deployed easily and rapidly\textcolor{blue}{\cite{9119122}}.

Owing to the above advantages, RIS-assisted wireless systems have drawn significant research interest recently, and many research works have been done, including high-layer protocol design\textcolor{blue}{\cite{9082859,9013979}}, channel estimation\textcolor{blue}{\cite{9039554, 9130088, 9133142}}, received signal-to-noise ratio (SNR) maximization\textcolor{blue}{\cite{8647620, 8811733, 8968350, 8683145}}, channel capacity analysis\textcolor{blue}{\cite{9148781, 8981888}}, energy efficiency maximization\textcolor{blue}{\cite{8644519, 8741198, Liu2012:Energy}}, quality-of-service (QoS) constrained transmit power minimization\textcolor{blue}{\cite{8930608, 8811733}}, and physical layer security design\textcolor{blue}{\cite{9288742, 9133130}}. It is worthy noting that all the mentioned researches study the RIS-assisted wireless system with an assumption that the hardware operations are perfect without any impairment. However, both physical transceiver and RIS suffer from inevitable hardware impairments\textcolor{blue}{\cite{4586304, 1381434, 6891254, 8869792}}. Different from the traditional noise, the hardware impairments at the transceiver lead to the mismatch or distortion between the intended signal and the actual signal, and the hardware impairments at the RIS cause the phase noise since perfectly precise configuration of the reflection phases is infeasible. Several existing works study the impact of hardware impairments on massive multiple-input multiple-output (MIMO) systems\textcolor{blue}{\cite{6362131, 6891254, 7835110}}. For example, it has been shown that hardware impairments limit the asymptotic channel capacity in massive MIMO systems and high-SNR slope thus collapsing to zero\textcolor{blue}{\cite{6891254}}.

The hardware impairments at transceivers and RIS, if not carefully handled, will affect both the direct link and the reflected link, which may lead to severer system performance degradation compared with conventional systems. Consequently, it is necessary to reevaluate the performance of RIS-assisted wireless systems with hardware impairments that includes channel estimation accuracy, uplink and downlink channel capacity, power scaling law, and energy efficiency. Furthermore, considering hardware impairments will also bring certain other significant challenges for fully utilizing RIS-assisted wireless systems. For example, in the case of perfect hardware, the optimal transmit beamforming solution is the maximum-ratio transmission (MRT), and the reflect beamforming optimization problem can be addressed by solving a non-convex quadratic constraint quadratic program (QCQP) problem\textcolor{blue}{\cite{8683145, 8930608, 8647620}}. Nevertheless, the consideration of hardware impairments increases the difficulty and complexity of the joint transmit and reflect beamforming optimization which is non-trivial even for the case of perfect hardware. As a result, it is necessary to reformulate the optimization problem and redesign the optimization algorithm.

Motivated by the above problems, by considering hardware impairments, we optimize the joint transmit and reflect beamforming and evaluate the system performance. To the best of our knowledge, it is the first work comprehensively evaluating the impact of hardware impairments on RIS-assisted wireless systems. The contributions of this paper are summarized as follows:
\begin{itemize}	
	\item By utilizing the novel three-phase pilot-based channel estimation framework, we give the closed-form estimators of the direct and cascaded channels in both single-user and multi-user cases with hardware impairments.
	Then, two corollaries about the impact of hardware impairments on estimation accuracy and how to reduce this impact are presented.

	\item We formulate an optimization problem with an aim to maximize the receive SNR for the RIS-assisted wireless system with hardware impairments. By transforming the optimization problem into a generalized Rayleigh quotient problem, we determine the optimal transmit beamforming and receive combining solutions.
	
	\item Based on the optimal transmit beamforming solution, a gradient descent method (GDM)-based algorithm is proposed to optimize the reflect beamforming.
	Our experimental results demonstrate that the proposed algorithm can offer a significant improvement in spectrum efficiency when the reflecting element number is appropriate. However, if we utilize massive reflecting elements, it is unnecessary to allocate resource to optimize the beamforming, and the channel capacity limit can still be reached which is caused by hardware impairments.
	
	\item By considering three types of asymptotics with respect to the transmit power, the BS antenna number, and the reflecting element number, we analyze the asymptotic channel capacities and derive the corresponding upper and lower bounds.
	
	\item In terms of the system energy consumption, we derive the power scaling law in both cases of perfect and imperfect channel state information (CSI) and further analyze the maximal energy efficiency with hardware impairments. Our experimental results also show that RIS-assisted wireless systems can achieve both high spectral efficiency and high energy efficiency without the need for massive BS antennas.
\end{itemize}

The rest of this paper is organized as follows. Section II presents the system model, and the hardware impairment model. The channel estimation performance is analyzed in Section III. In Section IV, we derive the optimal transmit beamforming solution and propose the GDM-based algorithm for the reflect beamforming optimization. The asymptotic downlink channel capacity is also discussed in Section IV. Section V presents the power scaling law and the maximal energy efficiency. Finally, we conclude the paper in Section VI. The numerical results are illustrated in each corresponding section.

\section{System and Hardware Impairment Models}
\begin{figure}[h]
	\centering
	\quad \quad \quad
	\includegraphics[width = 12.7 cm]{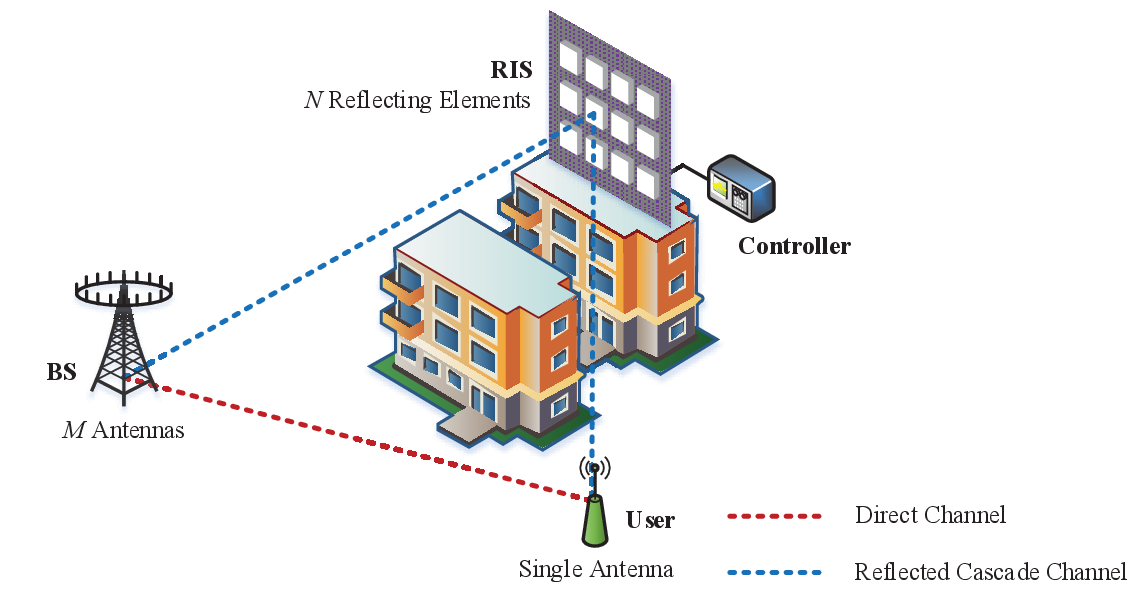}
	\caption{The RIS-assisted system with an \textit{M}-antenna BS, a single-antenna user, and an RIS comprising \textit{N} reflecting elements.}
	\label{fig1}
\end{figure}

In this section, we establish the system model based on the physically correct models in prior works. We consider an RIS-assisted wireless system where an RIS is deployed to assist the communication from a multi-antenna BS to a single-antenna user over a given frequency band, as illustrated in \textcolor{blue}{Fig.~\ref{fig1}}. 
\vspace{-13 pt}
\vspace{-3 pt}
The number of antennas at the BS and that of reflecting elements at the RIS are denoted by $ M $ and $ N $, respectively.  The operations at the RIS is represented by the diagonal matrix $\mathbf{\Phi} = \operatorname{diag}\left(e^{j\theta_{1}},\cdots,e^{j\theta_{N}}\right)$\footnote{
	Since the RIS is a passive device, each reflecting element is usually designed to maximize the signal reflection. Thus, we set the amplitude of reflection coefficient equal to one for simplicity in this paper.
}, where $\theta_{i} \in [0,2\pi)$ represents the phase shift\footnote{
	We assume that the phase shifts can be continuously varied in $ [0,2\pi) $, while they are usually selected from a finite number of discrete values in practice. Wu and Zhang address the beamforming design of RIS with discrete phase shifts, and show that the RIS with discrete phase shifts achieve the same power gain with that of the RIS with continuous phase shifts\textcolor{blue}{\cite{8930608}}. 
} of the $i$-th reflecting element and $j$ represents the imaginary unit. We assume that the channels follow Rayleigh block-fading, and the realizations of them are random and independent between blocks. Denote the channels of BS-user link, BS-RIS link, and RIS-user link as $ \mathbf{h}_{\mathrm{d}} \in \mathbb{C}^{M\times1} $, $ \mathbf{G} \in \mathbb{C}^{M \times N} $, and $ \mathbf{h}_{\mathrm{r}} \in \mathbb{C}^{N \times 1} $, respectively. They are modeled as ergodic processes with fixed independent realizations in each coherence period. The direct channel $ \mathbf{h}_{\mathrm{d}} \sim \mathcal{C}\mathcal{N} \left(\mathbf{0}, \mathbf{C}_{\mathrm{D}} \right)$ 
and the reflected cascade channel
$\mathbf{h}_{\mathrm{c}}=\mathbf{G}\mathbf{h}_{\mathrm{r}}\sim \mathcal{C}\mathcal{N}\left(\mathbf{0},\mathbf{C}_{\mathrm{R}}\right)$, where $\mathbf{C}_{\mathrm{D}} = \beta_{d} \mathbf{I}$ and $\mathbf{C}_{\mathrm{R}} = N\beta_{r} \mathbf{I}$. We also define the matrix $\mathbf{H}_{\mathrm{R}}=\mathbf{G}\operatorname{diag}\left(\mathbf{h}_{\mathrm{r}}\right)$, and then we have that $\mathbb{E}\{\mathbf{H}_{\mathrm{R}}\} = \mathbf{0}$ and $\mathbb{E}\{\mathbf{H}_{\mathrm{R}}\mathbf{H}_{\mathrm{R}}^{\mathrm{H}}\} = \mathbf{C}_{\mathrm{R}}$.

\subsection{Hardware Impairment Model}
The hardware impairments considered in this paper include transceiver hardware impairments and RIS hardware impairments. Although the physical transceiver implementations consist of many different hardware components and each one distorts the signals in its own way\textcolor{blue}{\cite{1532220}}, the aggregate residual hardware impairments can be modeled as independent additive distortion noise\textcolor{blue}{\cite{5456453, zetterberg2011experimental, 4641657, 6503874}}. Since the RIS is a passive device and high-precision configuration of the reflection phases is unfeasible, the RIS hardware impairments can be modeled as phase errors\textcolor{blue}{\cite{8869792}}. The detailed hardware impairment models can be established as follows.

\subsubsection{Transceiver Hardware Impairments}
The hardware impairments of transceiver cause the mismatch or distortion between the intended signal and the actual signal. The distortion can be well-modeled as uncorrelated additive Gaussian noise since it is the aggregate residual of many hardware impairments, where some are Gaussian and some behave as Gaussian when summed up\textcolor{blue}{\cite{5456453, zetterberg2011experimental}}. We use additive distortion noise $ \eta_{u} \in \mathbb{C} $ and $ \boldsymbol{\eta}_{b} \in \mathbb{C}^{M\times1} $ to describe the residual impairments of the transceiver hardwares at the BS and the user, respectively. They obey the circularly symmetric complex Gaussian distribution, $ \eta_{u} \sim \mathcal{C} \mathcal{N} \left(0, v_{u}\right) $ and $ \boldsymbol{\eta}_{b} \sim \mathcal{C N}\left(\mathbf{0}, \mathbf{\Upsilon}_{\mathrm{B}}\right) $, where $v_{u}$ and $\mathbf{\Upsilon}_{\mathrm{B}}$ are the variance and covariance matrix of the additive distortion noise, respectively. It should be noted that the distortion noise power at an antenna is proportional to the signal power at this antenna\textcolor{blue}{\cite{5456453, zetterberg2011experimental}}. When all the reflecting elements are switched off, there only exists the direct channel in the system. The variance $v_{u}$ of the additive distortion noise at the user can be modeled as $v_{u}=\kappa_{u} p_{u}$, and the covariance matrix $\mathbf{\Upsilon}_{\mathrm{B}}$ of the additive distortion noise at the BS can be modeled as $\mathbf{\Upsilon}_{\mathrm{B}}=\kappa_{b}\left(p_{u}+v_{u}\right) \mathbf{C}_{\mathrm{D}}$, where
$p_{u}$ is the transmit power, $\kappa_{u}$ and $\kappa_{b}$ are the proportionality coefficients which characterize the levels of hardware impairments at the user and the BS, respectively, and are related to the error vector magnitude (EVM). The EVM is a common measure of transceiver hardware quality\footnote{
More precisely, the EVM is the difference between the actual transmitted/received symbols and the ideal symbols. It can be degraded by various imperfections, e.g., noise, distortion, spurious signals, and phase noise. Thus, the EVM can provide a comprehensive measure of the quality of the radio transceiver for use in digital communications. The 3GPP LTE standard specifies total EVM requirements in the range $ [0.08, 0.175] $\textcolor{blue}{\cite{holma2011lte}}.
}, {e.g.}, when the BS transmits the signal $\boldsymbol{x}$, the EVM at the BS is defined as
\begin{equation}
\mathrm{EVM}_{\mathrm{BS}} = \sqrt{\frac{\operatorname{tr}\left( \mathbb{E}\left\{\boldsymbol{\eta}_{b} \boldsymbol{\eta}_{b}^{\mathrm{H}}\right\} \right) }{\operatorname{tr}\left( \mathbb{E}\left\{\boldsymbol{x} \boldsymbol{x}^{\mathrm{H}}\right\}\right) }} = \sqrt{\kappa_{b}} .
\end{equation}
When all the reflecting elements are switched on, there exist the reflected cascade channel besides the direct channel in the system. In the uplink, the variance $v_{u}$ of the additive distortion noise at the user is still $v_{u}=\kappa_{u} p_{u}$, 
while the covariance matrix $ \mathbf{\Upsilon}_{\mathrm{B}} $ of the additive distortion noise at the BS turns into $\mathbf{\Upsilon}_{\mathrm{B}}=\kappa_{b}\left(p_{u}+v_{u}\right)\mathbb{E}\{\mathbf{h}\mathbf{h}^{\mathrm{H}}\}$ where $\mathbf{h} = \mathbf{h}_{\mathrm{d}} + \mathbf{G\Phi} \mathbf{h}_{\mathrm{r}}$ is the overall channel vector. 
In the downlink, the covariance matrix $ \mathbf{\Upsilon}_{\mathrm{B}} $ of the additive distortion noise at the BS can be modeled as 
$\mathbf{\Upsilon}_{\mathrm{B}}=\kappa_{b} \operatorname{diag}\left(\mathbb{E}\{\boldsymbol{x}\boldsymbol{x}^{\mathrm{H}}\}\right)$, where $\boldsymbol{x}$ is the transmitted signal. The variance $v_{u}$ of the distortion noise at the user turns into 
$v_{u}=\kappa_{u} \mathbb{E} \{\mathbf{h}^{\mathrm{H}} (\mathbb{E}\{\boldsymbol{x}\boldsymbol{x}^{\mathrm{H}}\} + \mathbf{\Upsilon}_{\mathrm{B}}) \mathbf{h} \} $.

\subsubsection{RIS Hardware Impairments}
The hardware impairments of RIS can be modeled as phase noise since the RIS is a passive device and high-precision configuration of the reflection phases is infeasible. The phase noise at the $i$-th reflecting element is denoted as $\Delta \theta_{i}$, which is randomly distributed on $[-\pi, \pi]$ according to a certain circular distribution. According to the result in\textcolor{blue}{\cite{8869792}}, the distribution of the phase noise $\Delta \theta_{i}$ has mean direction zero, {i.e.}, $\arg (\mathbb{E}\{e^{j \Delta \theta_{i}}\})=0$, and its probability density function is symmetric around zero. Then, the actual response matrix of RIS with phase noise is $\hat{\mathbf{\Phi}}=\operatorname{diag}\left[e^{j(\theta_{1}+\Delta\theta_{1})}, e^{j(\theta_{2}+\Delta\theta_{2})}, \cdots, e^{j(\theta_{N}+\Delta\theta_{N})}\right]$.

\vspace{-10 pt}
\subsection{Received Signal Model}
Based on the communication system model given in the previous subsection, we establish the received signal model in this subsection. The uplink is used for pilot-based channel estimation and data transmission. The channel estimation strategy and the received pilot signal at the BS will be illustrated detailedly in the next section. The received data signal $\boldsymbol{y} \in\mathbb{C}^{M\times1}$ at the BS transmitted from the user can be expressed as follows:
\begin{equation}
\label{uplink}
\boldsymbol{y}
= \left(\mathbf{h}_{\mathrm{d}}+\mathbf{G} \hat{\mathbf{\Phi}} \mathbf{h}_{\mathrm{r}}\right)\left(x+\eta_{u}\right)+\boldsymbol{\eta}_{b}+\boldsymbol{n},
\end{equation}
where $x \in \mathbb{C}$ is the data signal transmitted by the user and the transmit power is $p_{u} = \mathbb{E}\{|x|^{2}\}$, and $\boldsymbol{n} \in \mathbb{C}^{M \times 1}$ is the additive white Gaussian noise (AWGN) with the elements independently drawn from $\mathcal{C} \mathcal{N}\left(0, \sigma_{b}^{2}\right)$. 
The downlink is used for data transmission, and the received data signal $y\in\mathbb{C}$ at the user from the BS can be expressed as follows:
\begin{equation}
\label{downlink}
y
=\left(\mathbf{h}_{\mathrm{d}}+\mathbf{G} \hat{\mathbf{\Phi}} \mathbf{h}_{\mathrm{r}}\right)^{\mathrm{H}} \left(\boldsymbol{x}+\boldsymbol{\eta}_{b}\right)+\eta_{u}+n,
\end{equation}
where $ \boldsymbol{x}\in\mathbb{C}^{M \times 1} $ is the data signal transmitted by the BS and $n\in\mathbb{C}$ is the AWGN drawn from $\mathcal{C}\mathcal{N}\left(0, \sigma_{u}^{2}\right)$. The transmit power of the BS is $ p_{b} = \mathbb{E}\{\boldsymbol{x}^{\mathrm{H}}\boldsymbol{x}\} $.

\section{Channel Estimation with Hardware Impairments }

In this section, we perform the channel estimations, including the estimations of direct channel and cascaded channel, for the considered system with hardware impairments. 
Several existing strategies have been proposed to estimate in the case without considering hardware impairments, e.g.,  an on-off state control based strategy is proposed in\textcolor{blue}{\cite{9039554}} and\textcolor{blue}{\cite{Liu2103:Channel}} to estimate each reflected cascade channel without interference from the other reflecting elements, and a novel hierarchical training reflection design is proposed to progressively estimate reflected cascade channels over multiple time blocks\textcolor{blue}{\cite{9133142}}. To reduce the channel estimation time, we adopt the novel three-phase pilot-based channel estimation framework proposed in\textcolor{blue}{\cite{9130088}} to make such estimations. In order to be comprehensive, we will evaluate the estimation performance affected by the existence of hardware impairments in both cases with single-user and multi-user. By utilizing this framework, the direct channel is first estimated for all users in the first phase, the cascade channel is then estimated for a typical user in the second phase, and finally we estimate the cascade channels for other users in the third phase. The theorem of LMMSE channel estimator and its MSE with hardware impairments and the two corollaries about the impact of hardware impairments on channel estimations will be illustrated in the later subsections.

\vspace{-5 pt}
\subsection{Single-user Case with Hardware Impairments}

For the case with single-user, only two phases are needed to make estimations which consist of $\tau_{1}$ and $\tau_{2}$ time slots, respectively. In the first phase, all the reflecting elements are switched off, and the BS estimates the direct channel. The user transmits the pilot $a^{(i)}$ to the BS at the $i$-th time slot, $i = 1, \dots, \tau_{1}$. Then, the received signals at the BS over the whole phase are
\begin{equation}
\label{singleUser-phaseI}
\mathbf{Y}_{\text{I}} 
= [\boldsymbol{y}^{(1)}, \cdots, \boldsymbol{y}^{(\tau_{1})}]
= \mathbf{h}_{\text{d}} \left( \boldsymbol{a}_{\text{I}}^{\mathrm{T}} + \boldsymbol{\eta}^{\mathrm{T}}_{u,\text{I}} \right) + [\boldsymbol{\eta}^{(1)}_{b}, \cdots, \boldsymbol{\eta}^{(\tau_{1})}_{b}] + [\boldsymbol{n}^{(1)}, \cdots, \boldsymbol{n}^{(\tau_{1})}],
\end{equation}
where $\boldsymbol{a}_{\text{I}}^{\mathrm{T}} = [a^{(1)}, \cdots, a^{(\tau_{1})}]$ and $\boldsymbol{\eta}^{\mathrm{T}}_{u,\text{I}} = [\eta^{(1)}_{u}, \cdots, \eta^{(\tau_{1})}_{u}]$.
Without loss of generality, we use the LMMSE estimator to make the channel estimation. From Eq. (\ref{singleUser-phaseI}), the MSE for estimating the direct channel $\mathbf{h}_{\mathrm{d}}$ can be expressed as
\begin{equation}
\begin{aligned}
\varepsilon_{\mathrm{I}}
& = \mathbb{E} \left[ (\hat{\mathbf{h}}_{\text{d}} - \mathbf{h}_{\text{d}})^{\mathrm{H}} (\hat{\mathbf{h}}_{\text{d}} - \mathbf{h}_{\text{d}}) \right] = \mathbb{E} \left[ \left(\mathbf{Y}_{\text{I}}\mathbf{A} - \mathbf{h}_{\text{d}}\right)^{\mathrm{H}} \left(\mathbf{Y}_{\text{I}}\mathbf{A} - \mathbf{h}_{\text{d}}\right) \right] \\ & = \mathbf{A}^{\mathrm{H}} \mathbb{E} \left(\mathbf{Y}_{\text{I}}^{\mathrm{H}} \mathbf{Y}_{\text{I}}\right) \mathbf{A} - \mathbb{E} \left(\mathbf{h}_{\text{d}}^{\mathrm{H}} \mathbf{Y}_{\text{I}}\right) \mathbf{A} - \mathbf{A}^{\mathrm{H}} \mathbb{E} \left(\mathbf{Y}_{\text{I}}^{\mathrm{H}} \mathbf{h}_{\text{d}}\right) + \mathbb{E} \left(\mathbf{h}_{\text{d}}^{\mathrm{H}} \mathbf{h}_{\text{d}}\right),
\end{aligned}
\end{equation}
where $\mathbf{A}$ is the detector matrix which can be obtained by equaling the partial derivative of $\varepsilon_{\mathrm{I}}$ with respect to $\mathbf{A}$  to zero, i.e., 
\begin{equation}
\frac{\partial \varepsilon_{\mathrm{I}}}{\partial \mathbf{A}} = \left[\mathbf{A}^{\mathrm{H}} \mathbb{E} \left(\mathbf{Y}_{\text{I}}^{\mathrm{H}} \mathbf{Y}_{\text{I}}\right)\right]^{\mathrm{T}} - \left[\mathbb{E} \left(\mathbf{h}_{\text{d}}^{\mathrm{H}} \mathbf{Y}_{\text{I}}\right)\right]^{\mathrm{T}} = \mathbf{0} \;\; \Rightarrow \; \mathbf{A} = \left[\mathbb{E} \left(\mathbf{Y}_{\text{I}}^{\mathrm{H}} \mathbf{Y}_{\text{I}}\right)\right]^{-1}  \left[\mathbb{E} \left(\mathbf{h}_{\text{d}}^{\mathrm{H}} \mathbf{Y}_{\text{I}}\right)\right]^{\mathrm{H}}.
\end{equation}
Therefore, the LMMSE estimator for direct channel and its MSE respectively are 
\begin{equation}
\label{single d estimator}
\hat{\mathbf{h}}_{\text{d}} = \frac{\beta_{d} \mathbf{Y}_{\text{I}} \; [a^{(1)}, \cdots, a^{(\tau_{1})}]^{\mathrm{H}}}{\beta_{d}\left( \tau_{1} p_{u} + v_{u}\right) + v_{b} + \sigma^{2}_{b}} ,
\end{equation}
\begin{equation}
\label{single d MSE}
\varepsilon_{\text{I}} = M\beta_{d} - \frac{\tau_{1} p_{u} M \beta^{2}_{d}}{\beta_{d}\left( \tau_{1} p_{u}+v_{u}\right) + v_{b} + \sigma^{2}_{b}} .
\end{equation}

In the second phase, all the reflecting elements are switched on such that the BS can estimate the reflected cascade channel. The user transmits the pilot $a^{(i)} = \sqrt{p_{u}}\;$ to the BS at the $i$-th time slot, $i = \tau_{1}+1, \dots, \tau_{1}+\tau_{2}$. Then, by canceling the interference caused by direct channel that is estimated in the first phase, the received signals at the BS over the whole phase are
\begin{equation}
\label{Y in Phase II}
\begin{aligned}
\tilde{\mathbf{Y}}_{\mathrm{II}} 
= & \left[\boldsymbol{y}^{(\tau_{1}+1)}, \cdots, \boldsymbol{y}^{(\tau_{1}+\tau_{2})}\right] - \hat{\mathbf{h}}_{\mathrm{d}} \boldsymbol{a}^{\mathrm{T}}_{\text{II}} \\ 
= & \; \mathbf{H} \hat{\mathbf{\Phi}}_{\text{II}} \left[\operatorname{diag} \left(\boldsymbol{a}^{\mathrm{T}}_{\text{II}} + \boldsymbol{\eta}^{\mathrm{T}}_{u,\text{II}}\right)\right] + {\mathbf{h}}_{\text{d}} \boldsymbol{\eta}^{\mathrm{T}}_{u,\text{II}} + (\mathbf{h}_{\mathrm{d}} - \hat{\mathbf{h}}_{\mathrm{d}}) \boldsymbol{a}^{\mathrm{T}}_{\text{II}} +\left[\boldsymbol{\eta}^{(\tau_{1}+1)}_{b}, \cdots, \boldsymbol{\eta}^{(\tau_{1}+\tau_{2})}_{b}\right] + \mathbf{N}_{\text{II}} ,
\end{aligned}	
\end{equation}
where $\boldsymbol{a}_{\text{II}}^{\mathrm{T}} = [a^{(\tau_{1}+1)}, \cdots, a^{(\tau_{1}+\tau_{2})}]$, $\boldsymbol{\eta}^{\mathrm{T}}_{u,\text{II}} = [\eta_{u}^{(\tau_{1}+1)}, \cdots, \eta^{(\tau_{1}+\tau_{2})}_{u}]$, $\mathbf{N}_{\text{II}} = \left[\boldsymbol{n}^{(\tau_{1}+1)}, \cdots, \boldsymbol{n}^{(\tau_{1}+\tau_{2})}\right]$ and $\hat{\mathbf{\Phi}}_{\mathrm{II}}$ is the $N \times \tau_{2}$ RIS matrix with the $(m,n)$-th entry equals $\operatorname{exp}[j(\theta_{m}^{(\tau_{1}+n)} + \Delta \theta_{m}^{(\tau_{1}+n)})]$. 
To ensure that $\operatorname{rank}(\hat{\mathbf{\Phi}}_{\mathrm{II}}) = N$, we construct ${\mathbf{\Phi}}_{\mathrm{II}}$ based on the discrete Fourier transform (DFT) matrix. For simplicity, we define $\dot{\mathbf{C}}_{\mathrm{R}} = \mathbb{E} \lbrace \left[\mathbf{h}_{1}, \cdots, \mathbf{h}_{N}\right]^{\mathrm{H}} \left[\mathbf{h}_{1}, \cdots, \mathbf{h}_{N}\right] \rbrace$ and $\mathbf{Z}_{\mathrm{II}}$ as the overall noise consisting of direct channel estimation error, hardware impairments and AWGN in Eq. (\ref{Y in Phase II}). Then, the covariance matrix of $\mathbf{Z}_{\mathrm{II}}$ can be expressed as
\begin{equation}
\label{12}
\mathbf{\Psi}_{\mathrm{II}} 
= \mathbb{E} \left[\mathbf{Z}_{\mathrm{II}}^{\mathrm{H}} \mathbf{Z}_{\mathrm{II}} \right] 
= v_{u} \operatorname{diag} \left( {\mathbf{\Phi}}_{\mathrm{II}}^{\mathrm{H}} \dot{\mathbf{C}}_{\mathrm{R}} {\mathbf{\Phi}}_{\mathrm{II}} \right)  
+ \left(v_{u} M \beta_{d} + M v_{b} + \tau_{2}p_{u} \varepsilon_{\mathrm{I}} + M \sigma^{2}_{b} \right)  \mathbf{I} .
\end{equation}
In this case, the LMMSE estimator for cascade channel and its MSE respectively are
\begin{equation}
\label{single R estimator}
\hat{\mathbf{H}}= \left[\hat{\mathbf{h}}_{1}, \cdots, \hat{\mathbf{h}}_{N}\right] 
= \sqrt{p_{u}}\; \tilde{\mathbf{Y}}_{\mathrm{II}} \left(p_{u}{\mathbf{\Phi}}_{\mathrm{II}}^{\mathrm{H}} \dot{\mathbf{C}}_{\mathrm{R}} {\mathbf{\Phi}}_{\mathrm{II}} + \mathbf{\Psi}_{\mathrm{II}}\right) ^{-1} {\mathbf{\Phi}}_{\mathrm{II}}^{\mathrm{H}} \dot{\mathbf{C}}_{\mathrm{R}} ,
\end{equation}
\begin{equation}
\label{single R MSE}
\varepsilon_{\text{II}} 
= \operatorname{Tr} \left[\dot{\mathbf{C}}_{\mathrm{R}} - p_{u} \dot{\mathbf{C}}_{\mathrm{R}} {\mathbf{\Phi}}_{\mathrm{II}} \left(p_{u}{\mathbf{\Phi}}_{\mathrm{II}}^{\mathrm{H}} \dot{\mathbf{C}}_{\mathrm{R}} {\mathbf{\Phi}}_{\mathrm{II}} + \mathbf{\Psi}_{\mathrm{II}}\right) ^{-1} {\mathbf{\Phi}}_{\mathrm{II}}^{\mathrm{H}} \dot{\mathbf{C}}_{\mathrm{R}} \right].
\end{equation}

\subsection{Multi-user Case with Hardware Impairments}

For the case with multi-user, there are three phases that consist of $\tau_{1}$, $\tau_{2}$ and $\tau_{3}$ time slots, respectively. In the first phase, all the reflecting elements are switched off, and the BS estimates the direct channels of all users $\mathbf{h}_{k,\mathrm{d}}$, $k \in \left\lbrace1, 2, \dots, K \right\rbrace$.
Each user $k$ transmits the pilot $a_{k}^{(i)}$ to the BS at the $i$-th time slot, $i = 1, \dots, \tau_{1}$. The pilot sequence of different users are set as orthogonal to each other, i.e., 
\begin{equation}
[\boldsymbol{a}_{1}, \cdots, \boldsymbol{a}_{K}]^{\mathrm{T}}  [\boldsymbol{a}_{1}, \cdots, \boldsymbol{a}_{K}]^{\mathrm{*}} = \tau_{1}
p_{u} \mathbf{I} .
\end{equation}
Then, the received signals at the BS over the whole phase can be expressed as
\begin{equation}
\begin{aligned}
\mathbf{Y}_{\mathrm{I}} = [\boldsymbol{y}^{(1)}, \cdots, \boldsymbol{y}^{(\tau_{1})}] 
 = &\left[ \mathbf{h}_{1,\mathrm{d}}, \dots, \mathbf{h}_{K,\mathrm{d}} \right]  [\boldsymbol{a}_{1}, \cdots, \boldsymbol{a}_{K}]^{\mathrm{T}} + [\boldsymbol{\eta}^{(1)}_{b}, \cdots, \boldsymbol{\eta}^{(\tau_{1})}_{b}] \\ 
 + &\left[ \mathbf{h}_{1,\mathrm{d}}, \dots, \mathbf{h}_{K,\mathrm{d}} \right] [\boldsymbol{\eta}_{1}, \cdots, \boldsymbol{\eta}_{K}]^{\mathrm{T}} + [\boldsymbol{n}^{(1)}, \cdots, \boldsymbol{n}^{(\tau_{1})}]  .
\end{aligned}	
\end{equation}
In this case, the LMMSE estimator for direct channel and its MSE, respectively, are 
\begin{equation}
\label{multi d estimator}
\left[ \hat{\mathbf{h}}_{1,\mathrm{d}}, \dots, \hat{\mathbf{h}}_{K,\text{d}} \right] 
= \frac{\beta_{d} \mathbf{Y}_{\mathrm{I}} [\boldsymbol{a}_{1}, \cdots, \boldsymbol{a}_{K}]^{\mathrm{*}}}{\beta_{d} \tau_{1} \left(p_{u} + v_{u}\right) + v_{b} + \sigma^{2}_{b}} ,
\end{equation}
\begin{equation}
\label{multi d MSE}
\varepsilon_{\text{I}} = K \left[M\beta_{d} - \frac{\tau_{1} p_{u} M \beta^{2}_{d}}{\beta_{d} \tau_{1} \left( p_{u}+v_{u}\right) + v_{b} + \sigma^{2}_{b}}\right] .
\end{equation}

In the second phase, all the reflecting elements are switched on, but only user 1 is permitted to transmit the pilot symbols to the BS. Then, the BS estimates the reflected cascade channel $\left[\mathbf{h}_{1,1}, \cdots, \mathbf{h}_{1,N}\right]$ of this user. This phase is same with that of the single-user case. 

In the third phase, all the users except user 1 transmit the pilot symbols to the BS. There exist relationships between the reflected cascade channel of user 1 and that of the other users:	 
\begin{equation}
\label{channel correlations}
\mathbf{h}_{k,n} = \frac{\left[\mathbf{h}_{k,\mathrm{r}}\right]_{n}}{\left[\mathbf{h}_{1,\mathrm{r}}\right]_{n}} \mathbf{h}_{1,n} = \lambda_{k,n} \mathbf{h}_{1,n}, \; k = 2, \dots, K; \; n = 1, \dots, N .
\end{equation}	 
By exploiting the channel correlations in Eq. (\ref{channel correlations}), the BS only needs to estimate the $(K-1)N$ parameters $\lambda_{2,n}, \cdots, \lambda_{K,n}$, $n = 1, \dots, N$. With the imperfect estimations of $\mathbf{h}_{k,\mathrm{d}}$ and $\mathbf{h}_{1,n}$, and canceling the interference caused by direct channels, the received signal at the $i$-th time slot, $i = \tau_{2}+1, \dots, \tau_{3}$, is given as follows:
\begin{equation}
\label{received signal PhaseIII}
\begin{aligned}
& \;\tilde{\boldsymbol{y}}^{(i)} = \boldsymbol{y}^{(i)} - \sum_{k=2}^{K} a^{(i)}_{k} \hat{\mathbf{h}}_{k,\text{d}} \\ 
& = \underbrace{\sum_{k=2}^{K} \sum_{n=1}^{N} a^{(i)}_{k} e^{j \left( \theta^{(i)}_{n} + \Delta \theta^{(i)}_{n} \right)} \lambda_{k,n} \hat{\mathbf{h}}_{1,n}}_{\text{Desired Signals}}  
+ \underbrace{\sum_{k=2}^{K} \eta^{(i)}_{u} \hat{\mathbf{h}}_{k,\text{d}} + \sum_{k=2}^{K} \sum_{n=1}^{N} \eta^{(i)}_{u} e^{j \left( \theta^{(i)}_{n} + \Delta \theta^{(i)}_{n} \right) }  \lambda_{k,n} \hat{\mathbf{h}}_{1,n} + \boldsymbol{\eta}_{b}^{(i)} }_{\text{Hardware Impairments}}  \\ 
& + \; \underbrace{\sum_{k=2}^{K} \left(a^{(i)}_{k} + \eta^{(i)}_{u} \right) \left( {\mathbf{h}}_{k,\text{d}} - \hat{\mathbf{h}}_{k,\mathrm{d}}\right)  + \sum_{k=2}^{K} \sum_{n=1}^{N} \left(a^{(i)}_{k} + \eta^{(i)}_{u} \right) e^{j \left( \theta^{(i)}_{n} + \Delta \theta^{(i)}_{n} \right) } \lambda_{k,n} \left(\mathbf{h}_{1,n}-\hat{\mathbf{h}}_{1,n}\right)}_{\text{Estimation Errors}} +\; \boldsymbol{n}^{(i)} .
\end{aligned}
\end{equation}	 
Furthermore, we utilize the orthogonal transmission and reflection strategy\textcolor{blue}{\cite{9130088}}. Specifically, at each time slot $i$, only one user $k_{i}$ is permitted to transmit pilot symbol, and only $M$ reflecting elements, denoted by set $\Delta_{i}$, are switched on, such that the BS is able to estimate the reflected cascade channels $\mathbf{h}_{k_{i}, n}$, $n \in \Delta_{i}$. As a result, for user $k_{i}$, $\lceil N/M \rceil$ time slots are allocated to estimate its reflected cascade channels. Then, the received signal in Eq. (\ref{received signal PhaseIII}) reduces to 
\begin{equation}
\label{signalofphaseIII}
\begin{aligned}
\tilde{\boldsymbol{y}}^{(i)} & = \underbrace{\sum_{n\in\Delta_{i}} a^{(i)}_{k_{i}} e^{j \left( \theta^{(i)}_{n} + \Delta \theta^{(i)}_{n} \right)} \lambda_{k_{i},n} \hat{\mathbf{h}}_{1,n}}_{\text{Desired Signals}}  
+ \underbrace{\eta^{(i)}_{u} \hat{\mathbf{h}}_{k_{i},\text{d}} + \sum_{n\in\Delta_{i}} \eta^{(i)}_{u} e^{j \left( \theta^{(i)}_{n} + \Delta \theta^{(i)}_{n} \right)} \lambda_{k_{i},n} \hat{\mathbf{h}}_{1,n} + \boldsymbol{\eta}_{b}^{(i)} }_{\text{Hardware Impairments}}  \\ 
& + \underbrace{\left(a^{(i)}_{k_{i}} + \eta^{(i)}_{u} \right) \left( {\mathbf{h}}_{k_{i},\text{d}} - \hat{\mathbf{h}}_{k_{i},\text{d}}\right)  + \sum_{n\in\Delta_{i}} \left(a^{(i)}_{k_{i}} + \eta^{(i)}_{u} \right) e^{j \left( \theta^{(i)}_{n} + \Delta \theta^{(i)}_{n} \right)} \lambda_{k_{i},n} \left(\mathbf{h}_{1,n}-\hat{\mathbf{h}}_{1,n}\right)}_{\text{Estimation Errors}} +\; \boldsymbol{n}^{(i)} .
\end{aligned}
\end{equation}
For convenience, we denote $\boldsymbol{z}^{(i)}_{\text{III}}$ as the overall noise at the $i$-th time slot, consisting of hardware impairments, estimation errors and AWGN in Eq. (\ref{signalofphaseIII}). We set the RIS phase shifts as zero. Then, the received signal at the $i$-th time slot can be simplified as
\begin{equation}
\tilde{\boldsymbol{y}}^{(i)} = a^{(i)}_{k_{i}} \mathbf{G}_{1,i} \boldsymbol{\lambda}_{k_{i},i} + \boldsymbol{z}^{(i)}_{\text{III}} ,
\end{equation}
where 
\begin{equation}
\quad \boldsymbol{\lambda}_{k_{i},i} = [\lambda_{k_{i},\Delta_{i}(1)}, \dots, \lambda_{k_{i},\Delta_{i}(|\Delta_{i}|)}]^{\mathrm{T}} ,
\end{equation}
\begin{equation}
\hat{\mathbf{G}}_{1,i} = [e^{j \Delta \theta^{(i)}_{\Delta_{i}(1)}} \hat{\mathbf{h}}_{1,\Delta_{i}(1)}, \dots, e^{j \Delta \theta^{(i)}_{\Delta_{i}(|\Delta_{i}|)}} \hat{\mathbf{h}}_{1,\Delta_{i}(|\Delta_{i}|)}] ,
\end{equation}
\begin{equation}
\label{phaseIIIerror}
\boldsymbol{z}^{(i)}_{\text{III}} =  
\eta^{(i)}_{u} \mathbf{G}_{1,i} \boldsymbol{\lambda}_{k_{i},i} + \boldsymbol{\eta}_{b}^{(i)} + \left(a^{(i)}_{k_{i}} + \eta^{(i)}_{u} \right) \left( {\mathbf{h}}_{k_{i},\text{d}} - \hat{\mathbf{h}}_{k_{i},\text{d}}\right) + \left(a^{(i)}_{k} + \eta^{(i)}_{u} \right) \left( \mathbf{G}_{1,i} - \hat{\mathbf{G}}_{1,i}\right)  \boldsymbol{\lambda}_{k_{i},i} + \boldsymbol{n}^{(i)} .
\end{equation}	
We observe that the estimation of channel $\mathbf{G}_{1,i}$ can be sufficiently accurate by increasing the pilot sequence length $\tau_{2}$ in the second phase. Thus, the estimation error term 
in Eq. (\ref{phaseIIIerror}) can be omitted. We define $\mathbf{C}_{k_{i},i} = \mathbb{E} [\boldsymbol{\lambda}_{k_{i},i} \boldsymbol{\lambda}_{k_{i},i}^{\mathrm{H}}] $. Then, the covariance matrix of $\boldsymbol{z}^{(i)}_{\text{III}}$ is 
\begin{equation}
\mathbf{\Psi}_{\mathrm{III}} 
= \mathbb{E} \left[\boldsymbol{z}^{(i)}_{\text{III}} \boldsymbol{z}^{(i)\mathrm{H}}_{\text{III}} \right] 
= v_{u} {\mathbf{G}}_{1,i} \mathbf{C}_{k_{i},i} {\mathbf{G}}_{1,i}^{\mathrm{H}} +\left(v_{b}+\sigma_{b}^{2} + \frac{\left(p_{u} + v_{u}\right) \varepsilon_{\text{I}}}{KM} \right) \mathbf{I} .
\end{equation}
The LMMSE channel estimator in the third phase and its MSE are respectively given by
\begin{equation}
\label{multi R estimator}
\hat{\boldsymbol{\lambda}}_{k_{i},i} = \sqrt{p_{u}} \mathbf{C}_{k_{i},i} \mathbf{G}_{1,i}^{\mathrm{H}} \left( p_{u}\mathbf{G}_{1,i} \mathbf{C}_{k_{i},i} \mathbf{G}_{1,i}^{\mathrm{H}} + \mathbf{\Psi}_{\mathrm{III}} \right)^{-1} \tilde{\boldsymbol{y}}^{(i)}  ,
\end{equation}
\begin{equation}
\label{multi R MSE}
\varepsilon_{\text{III}} = \operatorname{Tr} \left( \mathbf{C}_{k_{i},i} - p_{u} \mathbf{C}_{k_{i},i} \mathbf{G}_{1,i}^{\mathrm{H}} \left( p_{u}\mathbf{G}_{1,i} \mathbf{C}_{k_{i},i} \mathbf{G}_{1,i}^{\mathrm{H}} + \mathbf{\Psi}_{\mathrm{III}} \right)^{-1} \mathbf{G}_{1,i} \mathbf{C}_{k_{i},i}\right)  .
\end{equation}

\subsection{Impact of Hardware Impairments on Channel Estimations for RIS-assisted Systems}

In this subsection, we analyze the impact of hardware impairments on channel estimations for RIS-assisted systems, and present two corollaries about the channel estimation error floor and how to reduce the impact of hardware impairments. Before giving these two corollaries, we first summarize the previous two subsections, and give the following theorem:

\vspace{-5 pt}
\begin{theorem}{(LMMSE Channel Estimator and Its MSE with Hardware Impairments)}
	
\noindent By utilizing the three-phase based channel estimation framework\textcolor{blue}{\cite{9130088}}, the LMMSE estimators of direct channel(s) and cascaded channel(s) in an RIS-assisted system with hardware impairments in the cases of single-user and multi-user are given in Eqs. (\ref{single d estimator}), (\ref{single R estimator}), (\ref{multi d estimator}) and (\ref{multi R estimator}), respectively, and their corresponding MSEs are given in Eqs. (\ref{single d MSE}), (\ref{single R MSE}), (\ref{multi d MSE}) and (\ref{multi R MSE}).
\end{theorem}
\vspace{-5 pt}
\begin{IEEEproof}
Subsection A and B in this section constitute the proof of this theorem.
\end{IEEEproof}

\vspace{-5 pt}
\begin{corollary}{(Channel Estimation Error Floor Caused by Hardware Impairments)}
	
\noindent The estimation error per element in the direct channel is independent of the antenna number $M$ and reflecting element number $N$. The estimation error per element in the cascade channel is independent of the antenna number $M$, but depends on the reflecting element number $N$. The estimation error per element is a decreasing function of the pilot power, but does not converge to zero when the pilot power tends to infinity. The error floors of two phases in the case of single-user are respectively given as follows:
\begin{equation}
	\mu_{\mathrm{I}} = \beta_{d} - \frac{\tau_{1} \beta_{d}}{\tau_{1} + \kappa_{u} + \kappa_{b} + \kappa_{u}\kappa_{b}} ,
\end{equation}
\begin{equation}
\mu_{\mathrm{II}} = \beta_{r} - \frac{\tau_{2} \beta_{r}^2}{\left(\kappa_{u} + \kappa_{b} + \kappa_{u}\kappa_{b} \right) \left(\beta_{d} + N\beta_{r}\right) + \tau_{2}\mu_{\mathrm{I}} + \tau_{2} \beta_{r}}  .
\end{equation}

\end{corollary}

\begin{IEEEproof}
First, we consider the MSE of direct channel estimation in the first phase. In the high pilot power regime, the AWGN in the denominator of Eq. (\ref{single d MSE}) can be omitted, we have
\begin{equation}
\label{31}
\mu_{\mathrm{I}} 
= \lim_{p_{u} \rightarrow \infty} \frac{\varepsilon_{\text{I}}}{M} 
= \beta_{d} - \frac{\tau_{1} p_{u} \beta^{2}_{d}}{\beta_{d}\left( \tau_{1} p_{u}+v_{u}\right) + v_{b}}
= \beta_{d} - \frac{\tau_{1} \beta_{d}}{\tau_{1} + \kappa_{u} + \kappa_{b} + \kappa_{u}\kappa_{b}} .
\end{equation}
Then, we consider the MSE of cascade channel estimation in the second phase. 
By utilizing the matrix inversion lemma: \textcolor{black}{Woodbury formula}\footnote{
Let $\mathbf{A}$ be an $m \times m$ invertible matrix, $\mathbf{B}$ be an $n \times n$  invertible matrix, and $\mathbf{U}$ and $\mathbf{V}$ be $m \times n$ and $n \times m$ matrices, respectively, such that $(\mathbf{A} + \mathbf{U}\mathbf{B}\mathbf{V})$ is invertible, then it holds that $(\mathbf{A} + \mathbf{U}\mathbf{B}\mathbf{V})^{-1} = \mathbf{A}^{-1} - \mathbf{A}^{-1} \mathbf{U}\left(\mathbf{I} + \mathbf{B}\mathbf{V}\mathbf{A}^{-1}\mathbf{U}\right)^{-1} \mathbf{B} \mathbf{V} \mathbf{A}^{-1}$.
}\textcolor{blue}{\cite{matrices1950memorandum}}, the inverse matrix of  $( p_{u}{\mathbf{\Phi}}_{\mathrm{II}}^{\mathrm{H}} \dot{\mathbf{C}}_{\mathrm{R}} {\mathbf{\Phi}}_{\mathrm{II}} + \mathbf{\Psi}_{\mathrm{II}} )$ in Eq. (\ref{single R MSE}) can be expressed as follows:
\begin{equation}
\label{32}
\left(p_{u} {\mathbf{\Phi}}_{\mathrm{II}}^{\mathrm{H}} \dot{\mathbf{C}}_{\mathrm{R}} {\mathbf{\Phi}}_{\mathrm{II}} + \mathbf{\Psi}_{\mathrm{II}}\right)^{-1}
= \mathbf{\Psi}_{\mathrm{II}}^{-1} 
- p_{u} \mathbf{\Psi}_{\mathrm{II}}^{-1} {\mathbf{\Phi}}_{\mathrm{II}}^{\mathrm{H}} 
\left(\mathbf{I} 
+ p_{u} \dot{\mathbf{C}}_{\mathrm{R}} {\mathbf{\Phi}}_{\mathrm{II}} \mathbf{\Psi}_{\mathrm{II}}^{-1} {\mathbf{\Phi}}_{\mathrm{II}}^{\mathrm{H}} \right)^{-1}
\dot{\mathbf{C}}_{\mathrm{R}} {\mathbf{\Phi}}_{\mathrm{II}} \mathbf{\Psi}_{\mathrm{II}}^{-1} .
\end{equation}
Substituting Eq. (\ref{32}) into Eq. (\ref{single R MSE}) yields
\begin{equation}
\label{33}
\begin{aligned}
\varepsilon_{\text{II}} 
= & \operatorname{Tr} \left\lbrace \dot{\mathbf{C}}_{\mathrm{R}} - p_{u} \dot{\mathbf{C}}_{\mathrm{R}} {\mathbf{\Phi}}_{\mathrm{II}}  \mathbf{\Psi}_{\mathrm{II}}^{-1} {\mathbf{\Phi}}_{\mathrm{II}}^{\mathrm{H}} \dot{\mathbf{C}}_{\mathrm{R}} \right\rbrace \\
+ & \operatorname{Tr}\left\lbrace p_{u}^{2} \dot{\mathbf{C}}_{\mathrm{R}} {\mathbf{\Phi}}_{\mathrm{II}}  \mathbf{\Psi}_{\mathrm{II}}^{-1} {\mathbf{\Phi}}_{\mathrm{II}}^{\mathrm{H}} 
\left(\mathbf{I} + p_{u} \dot{\mathbf{C}}_{\mathrm{R}} {\mathbf{\Phi}}_{\mathrm{II}} \mathbf{\Psi}_{\mathrm{II}}^{-1} {\mathbf{\Phi}}_{\mathrm{II}}^{\mathrm{H}} \right)^{-1} \dot{\mathbf{C}}_{\mathrm{R}} {\mathbf{\Phi}}_{\mathrm{II}} \mathbf{\Psi}_{\mathrm{II}}^{-1} {\mathbf{\Phi}}_{\mathrm{II}}^{\mathrm{H}} \dot{\mathbf{C}}_{\mathrm{R}} \right\rbrace  .
\end{aligned}
\end{equation}
Since we set $\mathbf{\Phi}_{\text{II}}$ based on the DFT matrix, it holds that $[\mathbf{\Phi}_{\text{II}}]_{:,m} \times [\mathbf{\Phi}_{\text{II}}]_{:,m}^{\mathrm{H}} = N$. Thus, the first term $v_{u} \operatorname{diag} (\mathbf{\Phi}_{\text{II}}^{\mathrm{H}}\dot{\mathbf{C}}_{\mathrm{R}} \mathbf{\Phi}_{\text{II}} )$ of $\mathbf{\Psi}_{\mathrm{II}}$ in Eq. (\ref{12}) can be simplified as $v_{u} MN\beta_{r} \mathbf{I}$. Then, $\mathbf{\Psi}_{\mathrm{II}}$ is a diagonal matrix, and its inverse matrix can be easily obtained by
\begin{equation}
\label{34}
\mathbf{\Psi}_{\mathrm{II}}^{-1} = \frac{\frac{1}{M}}{v_{u} \left(\beta_{d} + N\beta_{r}\right) + v_{b} + \tau_{2}p_{u} \frac{\varepsilon_{\text{I}}}{M} + \sigma^{2}_{b}} \mathbf{I} \; \triangleq \; \psi \mathbf{I} .
\end{equation}
Since $\mathbf{\Phi}_{\text{II}}$ is constructed based on the DFT matrix and $\tau_{2} > N$, it follows that $\mathbf{\Phi}_{\text{II}}\mathbf{\Phi}_{\text{II}}^{\mathrm{H}} = \tau_{2} \mathbf{I}$. By substituting Eq. (\ref{34}) into Eq. (\ref{33}), the MSE $\varepsilon_{\text{II}}$ can be rewritten as
\begin{equation}
\label{35}
\begin{aligned}
\varepsilon_{\text{II}} = &\operatorname{Tr} \left\lbrace \dot{\mathbf{C}}_{\mathrm{R}} - p_{u} \tau_{2} \psi \dot{\mathbf{C}}_{\mathrm{R}} \dot{\mathbf{C}}_{\mathrm{R}}\right\rbrace \\ 
 + &\operatorname{Tr} \left\lbrace p_{u}^2 \tau_{2}^2 \psi^2 \dot{\mathbf{C}}_{\mathrm{R}} \left(\mathbf{I} + p_{u}\tau_{2}\psi  \dot{\mathbf{C}}_{\mathrm{R}} \right)^{-1} \dot{\mathbf{C}}_{\mathrm{R}} \dot{\mathbf{C}}_{\mathrm{R}} \right\rbrace \\
= &\;\frac{M N \beta_{r}}{ 1 + p_{u} \psi \tau_{2} M \beta_{r}} .
\end{aligned}
\end{equation}
When the pilot power tends to infinity, i.e., the transmit power $p_{u}$ tends to infinity, we have
\begin{equation}
\lim_{p_{u} \rightarrow \infty} p_{u} \psi 
= \frac{\frac{1}{M}}{\left(\kappa_{u} + \kappa_{b} + \kappa_{u}\kappa_{b} \right) \left(\beta_{d} + N\beta_{r}\right) + \tau_{2} \mu_{\text{I}}} .
\end{equation}
Substituting this limit into Eq. (\ref{35}) yields
\begin{equation}
\label{37}
\mu_{\text{II}} = \lim_{p_{u}\rightarrow \infty} \frac{\varepsilon_{\text{II}}}{MN} = \beta_{r} - \frac{\tau_{2} \beta_{r}^2}{\left(\kappa_{u} + \kappa_{b} + \kappa_{u}\kappa_{b} \right) \left(\beta_{d} + N\beta_{r}\right) + \tau_{2}\mu_{\text{I}} + \tau_{2} \beta_{r}} .
\end{equation}
Eqs. (\ref{31}) and (\ref{37}) are the non-zero estimation error floors and are independent of the antenna number $M$, while Eq. (\ref{37}) is dependent of the reflecting element number $N$.
\end{IEEEproof}

The extension of Corollary 1 to the scenario of multi-user is straightforward, and we omit it due to space limitation. Although perfect estimation accuracy cannot be achieved by increasing the pilot power due to the existence of hardware impairments, we can improve the MSE by increasing the pilot length, which is presented in the following corollary.

\vspace{-5 pt}
\begin{corollary}{(Reduce the Impact of Hardware Impairments by Increasing Pilot Length)}
	
\noindent Contrary to Corollary 1 that the MSE is limited by the non-zero error floor when we increase the pilot power, the MSE can be reduced to zero by increasing the pilot length although the hardware of RIS-assisted system is non-ideal.
\end{corollary}

\begin{IEEEproof}
First, we consider the MSE of direct channel estimation in the first phase. When the pilot length $\tau_{1}$ tends to infinity, the terms with no respect to $\tau_{1}$ in Eq. (\ref{single d MSE}) can be omitted, and we obtain the limit of $\varepsilon_{\text{I}}$ as follows: 
\begin{equation}
\label{38}
\lim_{\tau_{1} \rightarrow \infty} {\varepsilon_{\text{I}}} = M\beta_{d} - \frac{\tau_{1} p_{u} M \beta_{d}^2}{\tau_{1} p_{u} \beta_{d}} = 0 .
\end{equation}
Then, we consider the MSE of cascade channel estimation in the second phase. When the pilot lengths $\tau_{1}$ and $\tau_{2}$ tend to infinity, we have the following limit:
\begin{equation}
\lim_{\tau_{1},\tau_{2} \rightarrow \infty} \psi \tau_{2}  = \lim_{\tau_{2} \rightarrow \infty} \frac{\frac{1}{M}\tau_{2}}{v_{u} \left(\beta_{d} + N\beta_{r}\right) + v_{b} +  \frac{1}{M} \tau_{2} p_{u} \left( \mathop{\lim}\limits_{\tau_{1} \rightarrow \infty} \varepsilon_{\text{I}}\right)  + \sigma^{2}_{b}} = \frac{1}{p_{u}} \frac{1}{\mathop{\lim}\limits_{\tau_{1} \rightarrow \infty} \varepsilon_{\text{I}}}= \infty .
\end{equation}
Substituting this limit into Eq. (\ref{35}) yields
\begin{equation}
\label{40}
\lim_{\tau_{1},\tau_{2} \rightarrow \infty} {\varepsilon_{\text{II}}} = \frac{M N \beta_{r}}{ 1 +  p_{u} M \beta_{r}\left(\mathop{\lim}\limits_{\tau_{1},\tau_{2} \rightarrow \infty}\psi \tau_{2}\right)} = 0 .
\end{equation}
Eqs. (\ref{38}) and (\ref{40}) indicate that the MSE can be reduced to zero by increasing the pilot length although we have non-ideal hardwares, which proves the statement of this corollary.
\end{IEEEproof}

\subsection{Numerical Results}
In this subsection, we provide numerical results, which include theoretical characterizations and Monte Carlo simulations, to illustrate the channel estimation performance of the RIS-assisted wireless system with hardware impairments, and verify the validity of the two derived corollaries. 

We assume that the BS is equipped with $M = 10$ antennas and the RIS is equipped with $N = 25$ reflecting elements. The coefficients of impairment levels are chosen from the set of $\lbrace 0, 0.01^2, 0.05^2, 0.10^2\rbrace$. First, we show the estimation error per entry in the direct channel and the cascaded channel versus SNR which is defined as $\frac{p_{u}}{\sigma_{b}^2}$ in \textcolor{blue}{Fig.~2} and \textcolor{blue}{Fig.~3}, respectively. The theoretical characterizations match the Monte Carlo simulations perfectly. It is observed that the estimation error is a decreasing function of SNR or pilot power. It is also confirmed that there exist non-zero error floors at high SNRs, i.e., perfect channel estimation accuracy cannot be achieved by increasing the pilot power, which is proved in corollary 1. The error floor increases with the impairment level.
Moreover, by comparing \textcolor{blue}{Fig.~2} and \textcolor{blue}{Fig.~3}, we observe that the error floor of cascade channel estimation is slightly larger than that of direct channel estimation due to that the interference caused by direct channel cannot be completely canceled when we estimate the cascaded channel.
However, the error floor of cascaded channel estimation can be reached faster, i.e., only low pilot power is needed to fully utilize the RIS-assisted system.

\begin{figure}[htbp]
	\centering
	\begin{tabular}{cc}
		\begin{minipage}[t]{0.472\linewidth}
			\flushleft
			\includegraphics[width = 1\linewidth]{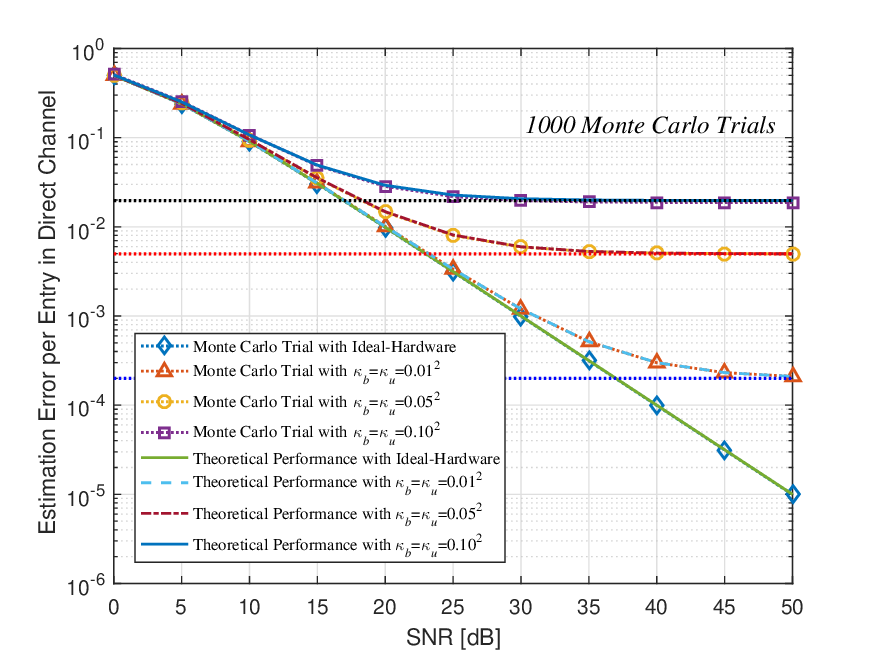}
			\vspace{-23 pt}
			\caption{The estimation error per entry in direct channel versus SNR with different impairment levels.}
		\end{minipage}
		\;\;
		\begin{minipage}[t]{0.472\linewidth}
			\includegraphics[width = 1\linewidth]{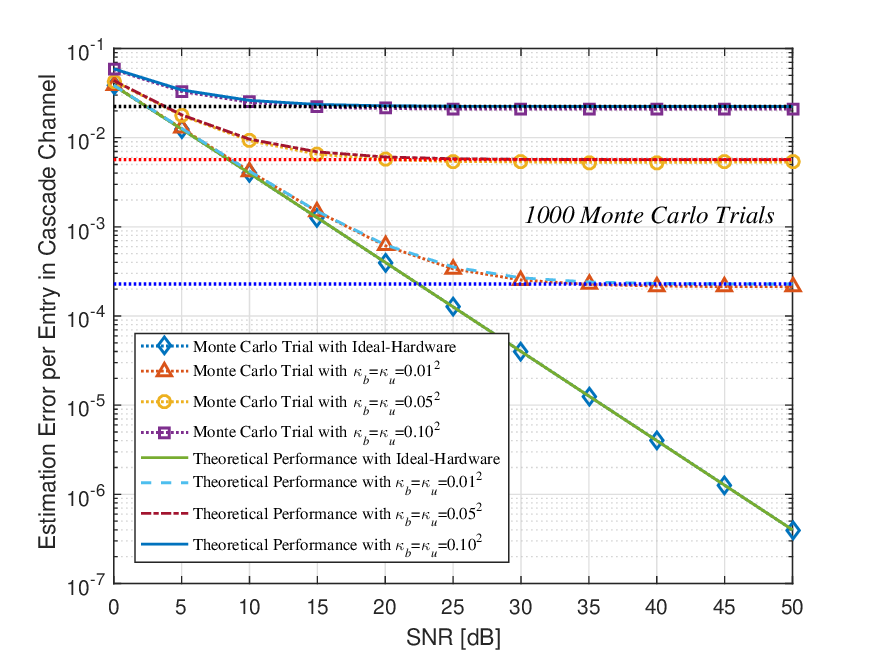}
			\vspace{-23 pt}
			\caption{The estimation error per entry in cascaded channel versus SNR with different impairment levels.}
		\end{minipage}
	\end{tabular}
\end{figure}

\vspace{-12 pt}

\begin{figure}[htbp]
	\centering
	\begin{tabular}{cc}
		\begin{minipage}[t]{0.472\linewidth}
			\flushleft
			\includegraphics[width = 1\linewidth]{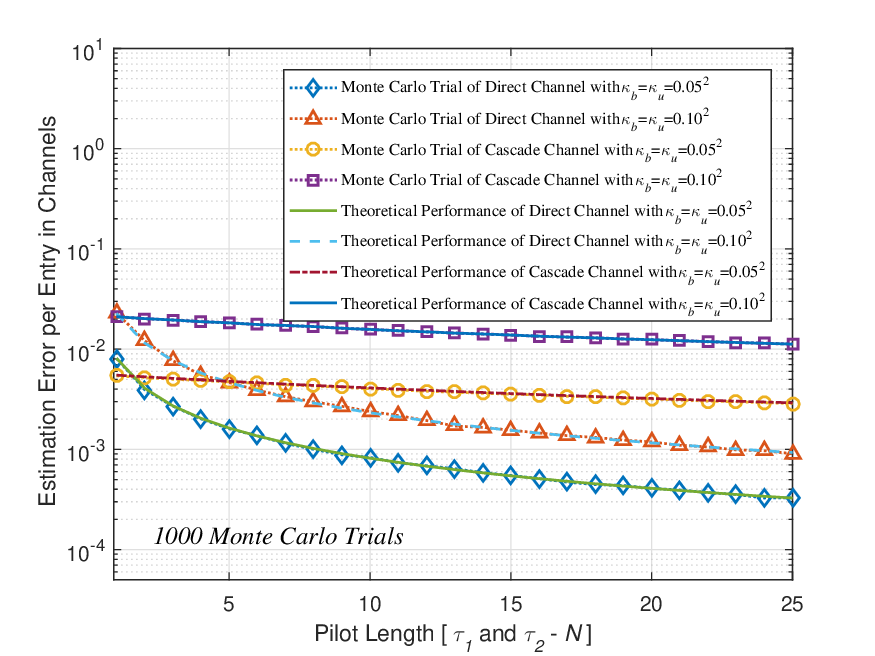}
			\vspace{-23 pt}
			\caption{The estimation error per entry in direct and cascaded channels versus pilot length with different impairment levels.}
		\end{minipage}
		\;\;
		\begin{minipage}[t]{0.472\linewidth}
			\includegraphics[width = 1\linewidth]{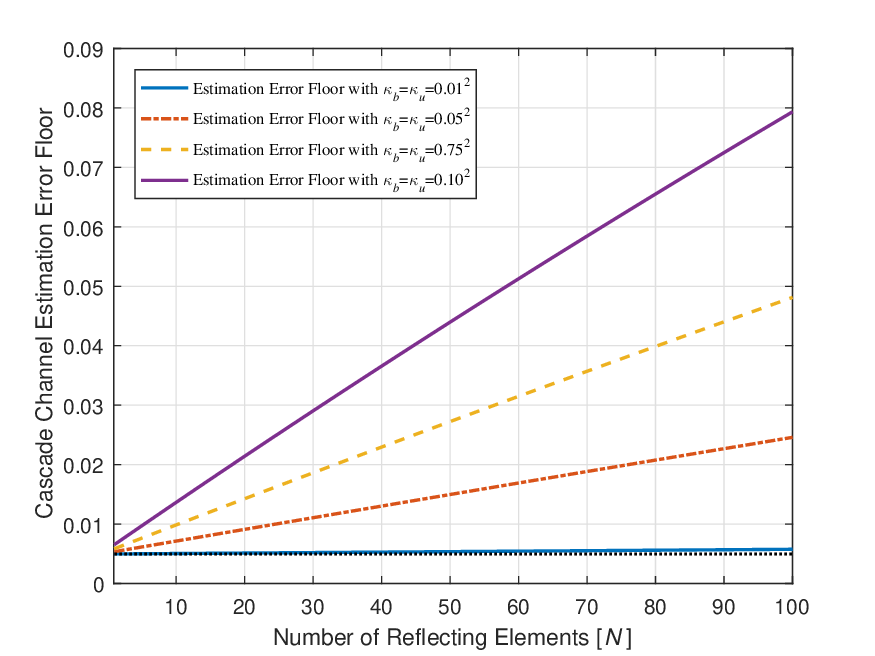}
			\vspace{-23 pt}
			\caption{The cascaded channel estimation error floor versus reflecting element number with different impairment levels.}
		\end{minipage}
	\end{tabular}
\end{figure}

\vspace{10 pt}

Next, we illustrate the improvement in estimation accuracy by increasing the pilot length in \textcolor{blue}{Fig.~4}. It is observed that the channel estimation error can be reduced to zero by increasing the pilot length although the hardware of RIS-assisted systems is non-ideal, which is proved in Corollary 2. The decline rate of direct channel estimation error is more rapid than that of cascaded channel estimation error when we increase the pilot length. In addition, we also give the relationship between the cascaded channel estimation error floor and the reflecting element number in \textcolor{blue}{Fig.~5}. It is observed that the error floor increases with the reflecting element number and the impairment level. This indicates that although using more reflecting elements can bring more spectral gains, it increases the cascaded channel estimation error.

\vspace{10 pt}

\section{Beamforming Design and Channel Capacity Analysis}

In this section, we optimize the joint (active) transmit and (passive) reflect beamforming, and derive the channel capacities of the uplink and downlink. In each coherence period $\tau$, by utilizing the channel estimation strategy analyzed in the previous section, the BS obtains an imperfect CSI $ \mathcal{H}_{b} $ of the actual channel state $ \mathcal{H} $, and the user obtains an imperfect CSI $ \mathcal{H}_{u} $. Based on the estimated CSI, the BS and the user select the conditional probability distribution of the transmitted data signal which we respectively denote as $ f(\boldsymbol{x}|\mathcal{H}_{b}) $ and $ f({x}|\mathcal{H}_{u}) $. Then, the channel capacities of the uplink and downlink can be expressed as
\begin{equation}
\label{C_up}
{{C}}_{u}=\frac{\tau_{u}}{\tau} \; \mathbb{E}\left\{\max \; \mathcal{I}\left(x ; \boldsymbol{y} | \mathcal{H}, \mathcal{H}_{b}, \mathcal{H}_{u}\right)\right\},
\end{equation}		
\begin{equation}
\label{C_down}
{{C}}_{d}=\frac{\tau_{d}}{\tau} \; \mathbb{E}\left\{\max \; \mathcal{I}\left(\boldsymbol{x} ; y | \mathcal{H}, \mathcal{H}_{b}, \mathcal{H}_{u}\right)\right\},
\end{equation}
where $ \mathcal{I}\left(x ; \boldsymbol{y} | \mathcal{H}, \mathcal{H}_{b}, \mathcal{H}_{u}\right) $ and $ \mathcal{I}\left(\boldsymbol{x} ; y | \mathcal{H}, \mathcal{H}_{b}, \mathcal{H}_{u}\right) $ are the conditional mutual information, and $\tau_{u}$ and $\tau_{d}$ are the uplink and downlink data transmission periods, respectively. The expectations in Eqs. (\ref{C_up}) and (\ref{C_down}) are taken over the joint distribution of $ \mathcal{H} $, $ \mathcal{H}_{b} $ and $ \mathcal{H}_{u} $. The maximizations in Eqs. (\ref{C_up}) and (\ref{C_down}) require a joint optimization of the active transmit (receive) beamforming (combining) at the BS and the passive reflect beamforming at the RIS.

The upper bounds on the uplink and downlink channel capacities in Eqs. (\ref{C_up}) and (\ref{C_down}) can be obtained by adding extra channel knowledge and removing all interference. Thus, we assume that the BS and the user have perfect channel state information, i.e., $ \mathcal{H}_{b}=\mathcal{H}_{u}=\mathcal{H} $. Since the receiver noise and the distortion noise caused by hardware impairments are circularly symmetric complex Gaussian distributed and independent of the useful signals under perfect CSI, Gaussian signaling is optimal in the uplink and downlink\textcolor{blue}{\cite{telatar1999capacity}}, and the single-stream transmission with $ \operatorname{rank}(\boldsymbol{x}\boldsymbol{x}^{\mathrm{H}})=1 $ is sufficient to achieve optimality\textcolor{blue}{\cite{6362131}}. We can set the transmitted data signal $ \boldsymbol{x} = \boldsymbol{w}_{b}^{d} x_{b}$ for $x_{b} \sim \mathcal{C} \mathcal{N}\left(0, p_{b}\right)$ at the BS and the transmitted data signal $ x \sim \mathcal{C} \mathcal{N}\left(0, p_{u}\right) $ at the user, where $\boldsymbol{w}_{b}^{d}$ is the unit-form beamforming vector\textcolor{blue}{\cite{6891254}}. 

To derive the channel capacity, we should optimize the transmit beamforming vector $ \boldsymbol{w}_{b}^{d} $ and the phase-shifting vector $ \boldsymbol{\theta}$. It is difficult due to the non-convex objective function as well as the unit-modulus constraints imposed by passive phase shifters. Although the existing researches on constant-envelope precoding\textcolor{blue}{\cite{6297982, 8047315}} and hybrid digital/analog processing\textcolor{blue}{\cite{6717211, 7389996}} have studied the beamforming optimizations under unit-modulus constraints, such designs are mainly restricted to the transceivers. The joint active and passive beamforming optimization has not been addressed. Moreover, considering hardware impairments increases the complexity of optimization, and no existing works have analyzed the reflect beamforming optimization under hardware impairments. To address this issue, we decouple the problem into two subproblems where we first  optimize the transmit beamforming vector $ \boldsymbol{w}_{b}^{d} $ to maximize the receive SNR for a given phase-shifting vector $ \boldsymbol{\theta} $, and then we optimize the phase-shifting vector $ \boldsymbol{\theta} $ based on the optimal transmit beamforming.

\subsection{Transmit Beamforming Optimization}
For any given phase shift at the RIS with ideal hardware, it can be verified that the maximum-ratio transmission (MRT) is the optimal solution\textcolor{blue}{\cite{8683145,8930608}}. Referring to\textcolor{blue}{\cite{6891254}}, this conclusion can be extended to the case with non-negligible hardware impairments, which is shown as follows.
\vspace{-5 pt}
\begin{theorem}
	\label{transmit beamforming}
	For any given phase shift at the RIS, the receive combining vector $ \boldsymbol{w}_{b}^{u} $ and the transmit beamforming vector $ \boldsymbol{w}_{b}^{d} $, which maximize the receive SNR in the RIS-assisted wireless communication system with hardware impairments, are respectively given by
	\begin{equation}
	\label{w_up}
	\boldsymbol{w}_{b}^{u}=\frac{\mathbf{U}^{-1} \hat{\mathbf{h}}^{\mathrm{H}}}{\; \left\|\mathbf{U}^{-1} \hat{\mathbf{h}}^{\mathrm{H}}\right\|_{2}} , \quad \boldsymbol{w}_{b}^{d}=\frac{\mathbf{D}^{-1} \hat{\mathbf{h}}}{\; \left\|\mathbf{D}^{-1} \hat{\mathbf{h}}\right\|_{2}}  ,
	\end{equation} 
	and the matrices $ \mathbf{U} $ and $ \mathbf{D} $ in Eq. (\ref{w_up}) are given by
	\begin{equation}
	\mathbf{U}=\left(1+\kappa_{u}\right) \kappa_{b} \operatorname{diag}(\hat{\mathbf{h}}\hat{\mathbf{h}}^{\mathrm{H}})+\kappa_{u} \hat{\mathbf{h}} \hat{\mathbf{h}}^{\mathrm{H}} + \frac{\sigma_{b}^{2}}{p_{u}} \mathbf{I},
	\end{equation}
	\begin{equation}
	\mathbf{D}=\left(1+\kappa_{u}\right) \kappa_{b} \operatorname{diag}(\hat{\mathbf{h}}\hat{\mathbf{h}}^{\mathrm{H}})+\kappa_{u} \hat{\mathbf{h}} \hat{\mathbf{h}}^{\mathrm{H}} + \frac{\sigma_{u}^{2}}{p_{b}} \mathbf{I},
	\end{equation}	
	where $ \hat{\mathbf{h}}$ represents the overall channel $( \mathbf{h}_{\mathrm{d}}+\mathbf{G}\hat{\mathbf{\Phi}}\mathbf{h}_{\mathrm{r}}) $.
\end{theorem}
\begin{IEEEproof}
	Suppose that the phase-shifting vector $ \boldsymbol{\theta} $ at the RIS is optimal, and then the upper bound of downlink channel capacity in Eq. (\ref{C_down}) can be written as
	\begin{equation}
	\label{C_down_SNR}
	{C}_{d} \leq \frac{\tau_{d}}{\tau} \; \mathbb{E}\left\{\max _{\|\boldsymbol{w}_{b}^{d}\|_{2}=1} \log _{2}(1+\operatorname{SNR}(\boldsymbol{w}_{b}^{d}))\right\},
	\end{equation}
	where
	\begin{equation}
	\label{SNR}
	\operatorname{SNR}(\boldsymbol{w}_{b}^{d})=\frac{|\hat{\mathbf{h}}^{\mathrm{H}}\boldsymbol{x}|^{2}} {\hat{\mathbf{h}}^{\mathrm{H}} \mathbf{\Upsilon}_{\mathrm{B}} \hat{\mathbf{h}} + v_{u} + \sigma_{u}^{2} } .
	\end{equation}
	According to the models of hardware impairments established in section II, the term $\hat{\mathbf{h}}^{\mathrm{H}} \mathbf{\Upsilon}_{\mathrm{B}} \hat{\mathbf{h}}$ in the denominator of Eq. (\ref{SNR}) can be written as
	\begin{equation}
	\label{term}
	\hat{\mathbf{h}}^{\mathrm{H}} \mathbf{\Upsilon}_{\mathrm{B}} \hat{\mathbf{h}} = \kappa_{b} \hat{\mathbf{h}}^{\mathrm{H}} \operatorname{diag}\left(\boldsymbol{x}\boldsymbol{x}^{\mathrm{H}}\right) \hat{\mathbf{h}} .
	\end{equation}
	Since the transmitted data signal at the BS has a form of $ \boldsymbol{x} = \boldsymbol{w}_{b}^{d} x_{b} $ and the variance of $ x_{b} $ is $p_{b}$, Eq. (\ref{term}) can be rewritten as 
	\begin{equation}
	\label{term1}
	\hat{\mathbf{h}}^{\mathrm{H}} \mathbf{\Upsilon}_{\mathrm{B}} \hat{\mathbf{h}} = \kappa_{b} p_{b} \left(\boldsymbol{w}_{b}^{d}\right)^{\mathrm{H}} \operatorname{diag}\left(\hat{\mathbf{h}}\hat{\mathbf{h}}^{\mathrm{H}}\right)\boldsymbol{w}_{b}^{d} .
	\end{equation}
	Similarly, the term $ |\hat{\mathbf{h}}^{\mathrm{H}}\boldsymbol{x}|^{2} $ in the numerator of Eq. (\ref{SNR}) and the term of $ v_{u} $ in the denominator of Eq. (\ref{SNR}) can be, respectively, rewritten as 
	\begin{equation}
	\label{numerator}
	|\hat{\mathbf{h}}^{\mathrm{H}}\boldsymbol{x}|^{2} = p_{b} \left(\boldsymbol{w}_{b}^{d}\right)^{\mathrm{H}} \hat{\mathbf{h}} \hat{\mathbf{h}}^{\mathrm{H}} \boldsymbol{w}_{b}^{d} ,
	\end{equation}
	\begin{equation}
	\label{term2}
	v_{u} = \kappa_{u} p_{b} \left(\boldsymbol{w}_{b}^{d}\right)^{\mathrm{H}} \hat{\mathbf{h}} \hat{\mathbf{h}}^{\mathrm{H}} \boldsymbol{w}_{b}^{d} + \kappa_{u} \hat{\mathbf{h}}^{\mathrm{H}} \mathbf{\Upsilon}_{\mathrm{B}} \hat{\mathbf{h}} .
	\end{equation}
	Then, by substituting Eqs. (\ref{term1}), (\ref{numerator}) and (\ref{term2}) into Eq. (\ref{SNR}), we obtain 
	\begin{equation}
	\label{121}
	\operatorname{SNR}(\boldsymbol{w}_{b}^{d})=\frac{\left(\boldsymbol{w}_{b}^{d}\right)^{\mathrm{H}} \hat{\mathbf{h}} \hat{\mathbf{h}}^{\mathrm{H}} \boldsymbol{w}_{b}^{d}\;} {\left(\boldsymbol{w}_{b}^{d}\right)^{\mathrm{H}} \mathbf{D} \boldsymbol{w}_{b}^{d}\;} ,
	\end{equation}
	where $\mathbf{D}=\left(1+\kappa_{u}\right) \kappa_{b} \operatorname{diag}(\hat{\mathbf{h}} \hat{\mathbf{h}}^{\mathrm{H}})+\kappa_{u} \hat{\mathbf{h}} \hat{\mathbf{h}}^{\mathrm{H}} + \frac{\sigma_{u}^{2}}{p_{b}} \mathbf{I}$.
	Function $ \log _{2}(1+\operatorname{SNR}(\boldsymbol{w}_{b}^{d}))$ in Eq. (\ref{C_down_SNR}) has a structure of $ f(x) = \log _{2}(1+x) $, and $ f(x) $ is a monotonically increasing function. Consequently, the maximum of Eq. (\ref{C_down_SNR}) can be obtained by maximizing the term $\operatorname{SNR}(\boldsymbol{w}_{b}^{d})$ in Eq. (\ref{121}), which is a generalized Rayleigh quotient problem. By solving this problem, the optimal transmit beamforming vector can be derived in Eq. (\ref{w_up}). The proof of the optimal receive combining vector follows the similar procedures and here we omit them due to space limitation.	
\end{IEEEproof}

Based on the optimal receive combining and transmit beamforming given in Theorem \ref{transmit beamforming}, we obtain the new upper bounds of the uplink and downlink channel capacities as follows:
\begin{equation}
{{C}}_{u} \leq \frac{\tau_{u}}{\tau} \; \mathbb{E}\left\{ \max_{\boldsymbol{\theta}} \; \log _{2}(1+\hat{\mathbf{h}}^{\mathrm{H}} \mathbf{U}^{-1} \hat{\mathbf{h}})\right\} ,
\end{equation}
\begin{equation}
\label{downlink channel capacity}
{{C}}_{d} \leq \frac{\tau_{d}}{\tau} \; \mathbb{E}\left\{ \max_{\boldsymbol{\theta}} \; \log _{2}(1+\hat{\mathbf{h}}^{\mathrm{H}} \mathbf{D}^{-1} \hat{\mathbf{h}})\right\} .
\end{equation}

\subsection{Reflect Beamforming Design}	
Based on the optimal transmit beamforming vector derived above, the reflect beamforming optimization can be analyzed in this subsection, which can be formulated as follows:
\begin{equation}
\label{p1}
\begin{aligned}
\text{(P1)}: \quad \max_{\boldsymbol{\theta}} &\quad \frac{\tau_{d}}{\tau} \log _{2}\left( 1+\hat{\mathbf{h}}^{\mathrm{H}} \mathbf{D}^{-1} \hat{\mathbf{h}}\right) \\
\operatorname{s.t.} &\quad {0 \leq \theta_{n} \leq 2 \pi, \quad \forall n=1, \cdots, N}, 
\end{aligned}
\end{equation}	
where $\mathbf{D}=\left(1+\kappa_{u}\right) \kappa_{b} \operatorname{diag}(\hat{\mathbf{h}}\hat{\mathbf{h}}^{\mathrm{H}})+\kappa_{u} \hat{\mathbf{h}} \hat{\mathbf{h}}^{\mathrm{H}} + \frac{\sigma_{u}^{2}}{p_{b}} \mathbf{I}$, and the optimization variable $ \boldsymbol{\theta} $ is hidden in the overall channel vector $ \hat{\mathbf{h}} $. By using Eq. (2.2) in\textcolor{blue}{\cite{silverstein1995empirical}}\footnote{
		For a complex scalar $ \tau \in \mathbb{C} $, a vector $ \boldsymbol{q} \in \mathbb{C}^{N} $, and a matrix $ \boldsymbol{B} \in \mathbb{C}^{N \times N} $,  where $ \boldsymbol{B} $ and $ ( \boldsymbol{B} + \tau\boldsymbol{q}\boldsymbol{q}^{\mathrm{H}}) $ are invertible, it holds that $ \boldsymbol{q}^{\mathrm{H}} (\boldsymbol{B} + \tau\boldsymbol{q}\boldsymbol{q}^{\mathrm{H}})^{-1} = {\boldsymbol{q}^{\mathrm{H}} \boldsymbol{B}^{-1}} / {(1+\tau\boldsymbol{q}^{\mathrm{H}}\boldsymbol{B}^{-1}\boldsymbol{q})} $.
}, problem (P1) can be transformed to the following equivalent problem:
\begin{equation}
\begin{aligned}
\text{(P2)}: \quad \max _{\boldsymbol{\theta}} &\quad \frac{\tau_{d}}{\tau} \log _{2}\left(1 + \frac{\hat{\mathbf{h}}^{\mathrm{H}} \tilde{\mathbf{D}}^{-1}\hat{\mathbf{h}}}{1+\kappa_{u}\hat{\mathbf{h}}^{\mathrm{H}} \tilde{\mathbf{D}}^{-1} \hat{\mathbf{h}}}\right) \\
\operatorname{s.t.} &\quad {0 \leq \theta_{n} \leq 2 \pi, \quad \forall n=1, \cdots, N},
\end{aligned}
\end{equation}	
where $ \tilde{\mathbf{D}}=\left(1+\kappa_{u}\right) \kappa_{b} \operatorname{diag}(\hat{\mathbf{h}}\hat{\mathbf{h}}^{\mathrm{H}})+ \frac{\sigma_{u}^{2}}{p_{b}} \mathbf{I}$. 
Problem (P2) has a structure of $ f(x) = \log _{2}(1+\frac{x}{1+ \alpha x}) $, $ \alpha > 0 $, and $ f(x) $ is a monotonically increasing function. Thus, problem (P2) can be optimized by maximizing $ \hat{\mathbf{h}}^{\mathrm{H}} \tilde{\mathbf{D}}^{-1}\hat{\mathbf{h}} $, which yields the following equivalent problem:
\begin{equation}
\label{p3}
\begin{aligned}
\text{(P3)}: \quad \max _{\boldsymbol{\theta}} &\quad \hat{\mathbf{h}}^{\mathrm{H}} \tilde{\mathbf{D}}^{-1} \hat{\mathbf{h}} \\
\operatorname{s.t.} &\quad {0 \leq \theta_{n} \leq 2 \pi, \quad \forall n=1, \cdots, N}.
\end{aligned}
\end{equation}	
According to the models of direct and reflected cascade channels established in section II, the overall channel vector $ \hat{\mathbf{h}}=\mathbf{h}_{\mathrm{d}}+\mathbf{G}\hat{\mathbf{\Phi}}\mathbf{h}_{\mathrm{r}} $ can be equivalently expressed as $ \hat{\mathbf{h}}=\mathbf{h}_{\mathrm{d}}+\mathbf{H}_{\mathrm{R}} \boldsymbol{v} $, where $ \boldsymbol{v}=[e^{j(\theta_{1}+\Delta\theta_{1})}, \cdots, e^{j(\theta_{N}+\Delta\theta_{N})}]^{\mathrm{T}} $. This formulation enables the separation of the response at the RIS in $ \boldsymbol{v} $ from the cascaded channel in $ \mathbf{H}_{\mathrm{R}} $. Since $ \tilde{\mathbf{D}} $ is a diagonal matrix, the inverse matrix of it can be easily obtained by
\begin{equation}
\tilde{\mathbf{D}}^{-1} = \operatorname{diag} \left(
\begin{array}{c}{\frac{1}{\left(1+\kappa_{u}\right) \kappa_{b} \left|{h}_{d, 1} + \boldsymbol{\varrho}^{\mathrm{T}}_{1} \boldsymbol{v}\right|^{2} + \frac{\sigma_{u}^{2}}{p_{b}}}}, \\ {\cdots} ,\\ {\frac{1}{\left(1+\kappa_{u}\right) \kappa_{b} \left|{h}_{d, M} + \boldsymbol{\varrho}^{\mathrm{T}}_{M} \boldsymbol{v}\right|^{2}+\frac{\sigma_{u}^{2}}{p_{b}}}}\end{array}
\right) ,
\end{equation}
where $ {h}_{d, i} $ is the $ i $-th entry of the direct channel vector $ \mathbf{h}_{\mathrm{d}} $, and  $\boldsymbol{\varrho}^{\mathrm{T}}_{i}$ is the $ i $-th row vector of the cascaded channel matrix $ \mathbf{H}_{\mathrm{R}} $. It should be noted that the phase errors of the reflecting elements are random and unknown to the BS, and the BS can only use the statistic characteristics of the phase errors, {i.e.}, $\arg \left(\mathbb{E}[e^{j \Delta \theta_{i}}]\right)=0$, to design the reflect beamforming, which yields $ \boldsymbol{v}=[e^{j\theta_{1}}, \cdots, e^{j\theta_{N}}]^{\mathrm{T}} $. In addition, the objective function is periodic with respect to $ \theta_{i}$ owing to the fact that $ e^{j(\theta_{i}+2k\pi)} = e^{j\theta_{i}} $ ($ k $ is an integer). Therefore, the constraints in problem (P3) can be omitted. Then, problem (P3) can be transformed to an unconstrained optimization problem:
\begin{equation}
\label{p4}
\text{(P4)}: \; \max _{\boldsymbol{\theta}} \;
\sum_{i=1}^{M} \frac{\left|{h}_{d, i}+\boldsymbol{\varrho}^{\mathrm{T}}_{i} \boldsymbol{v}\right|^{2}}{\left(1+\kappa_{u}\right)\kappa_{b}\left|{h}_{d, i} + \boldsymbol{\varrho}^{\mathrm{T}}_{i} \boldsymbol{v}\right|^{2}+\frac{\sigma_{u}^{2}}{p_{b}}}  .
\end{equation}	
We summarize the above result as the following proposition.

\vspace{-10 pt}
\begin{proposition}
	Based on the optimal transmit beamforming solution in Theorem \ref{transmit beamforming}, the reflect beamforming optimization, which is aimed to maximize the channel capacity of the RIS-assisted wireless communication system with hardware impairments, can be formulated as a non-convex optimization problem (P1), and can be simplified as an unconstrained optimization problem (P4).
\end{proposition}

\begin{remark}
	Problem (P4) has a different form with the reflect beamforming optimization problem solved by prior works where the problem is formulated as a non-convex quadratic constraint quadratic program (QCQP), e.g., [\textcolor{blue}{\citen{8930608}}, Eq. (4)], [\textcolor{blue}{\citen{8647620}}, Eq. (4)], and [\textcolor{blue}{\citen{8683145}}, Eq. (8)]. The main cause for the form difference is that we consider hardware impairments and make them as noise terms in SNR. These noise terms are not only  related to hardware quality, but also depend on the overall channel which contains the optimization variable. The consideration of hardware impairments increases the complexity of the optimization problem. This also implies that if we use cheaper and lower-quality hardware, {i.e.}, the hardware impairments are non-negligible in communication systems, the better reflect beamforming solution is needed to improve the channel quality. 
\end{remark}

Due to the objective function is a sum of the non-convex fractional function in problem (P4), it is difficult to obtain the global optimal solution. To further analyze this problem, we utilize the gradient decent method (GDM) to optimize the reflect beamforming. The gradient of objective function is a natural choice for the search direction\textcolor{blue}{\cite{boyd2004convex}}. The objective function of problem (P4) is a real-valued scalar function of the vector-variable. By introducing the conjugate coordinates $ \left( \boldsymbol{v}, \boldsymbol{v}^{*}\right) $, the objective function of problem (P4) can be guaranteed as a holomorphic function. The vector-variables $ \boldsymbol{v} $ and $ \boldsymbol{v}^{*} $ separately raise each element of $ [e^{j}, \cdots, e^{j}]^{\mathrm{T}} $ and $ [e^{-j}, \cdots, e^{-j}]^{\mathrm{T}} $ to the corresponding powers in $ \boldsymbol{\theta} $, {i.e.}, perform an element-wise power operation, which is according with IEEE Standard (754-2019) for floating-point arithmetic\textcolor{blue}{\cite{8766229}}. Thus, the objective function $ f\left( \boldsymbol{v}(\boldsymbol{\theta}), \boldsymbol{v}^{*}(\boldsymbol{\theta})\right)  $ is a composite function with the intermediate variables of $ \boldsymbol{v} $ and $ \boldsymbol{v}^{*} $, and the dependent variable (optimization variable) of $ \boldsymbol{\theta} $. According to the Jacobian matrix identification and the rules of gradient computation of a real-valued function with respect to its matrix variable\textcolor{blue}{\cite{zhang2017matrix,lutkepohl1996handbook}}, we obtain the gradients of the objective function with respect to $ \boldsymbol{v} $ and $ \boldsymbol{v}^{*} $, which are respectively shown as follows:
\begin{equation}
\frac{\partial f(\boldsymbol{v}, \boldsymbol{v}^{*})}{\partial \boldsymbol{v} } 
= \left[\frac{\partial f(\boldsymbol{v}, \boldsymbol{v}^{*})}{\partial \boldsymbol{v}^{*} }\right]^{*} = \frac{\sigma_{u}^{2}}{p_{b}} 
\sum_{i=1}^{M} 
\frac{h^{*}_{d, i} \boldsymbol{\varrho}_{i} +  \boldsymbol{\varrho}_{i} \boldsymbol{\varrho}_{i}^{\mathrm{H}} \boldsymbol{v}^{*}}
{ \left( \kappa \left| h_{d, i}+\boldsymbol{\varrho}_{i}^{\mathrm{T}} \boldsymbol{v}\right|^{2}+\frac{\sigma_{u}^{2}}{p_{b}}\right)^{2}} ,
\end{equation}
where $ \kappa = \left(1+\kappa_{u}\right)\kappa_{b} $. Similarly, the derivative of $ \boldsymbol{v} $ with respect to $ \boldsymbol{\theta} $, which we denote as $\mathbf{V}$, is given as follows:
\begin{equation}
\frac{\partial \boldsymbol{v}}{\partial \boldsymbol{\theta} } = \left[\frac{\partial \boldsymbol{v}^{*}}{\partial \boldsymbol{\theta} }\right]^{*} = \operatorname{diag}\left( je^{j\theta_{1}}, \cdots, je^{j\theta_{N}}\right).
\end{equation}
By using the derivation rule for composite function, we obtain the gradient of the objective function with respect to $ \boldsymbol{\theta} $,
\begin{equation}
\label{gradient}
\begin{aligned}
\nabla_{\boldsymbol{\theta}} f \left( \boldsymbol{v}(\boldsymbol{\theta}),\boldsymbol{v}^{*}(\boldsymbol{\theta})\right)   & = \frac{\partial f(\boldsymbol{v},\boldsymbol{v}^{*})}{\partial \boldsymbol{v} }  \frac{\partial \boldsymbol{v}}{\partial \boldsymbol{\theta}} + \frac{\partial f(\boldsymbol{v},\boldsymbol{v}^{*})}{\partial \boldsymbol{v}^{*} }  \frac{\partial \boldsymbol{v}^{*}}{\partial \boldsymbol{\theta} }  \\  
& = \frac{2\sigma_{u}^{2}}{p_{b}} \operatorname{Re} \left\lbrace \mathbf{V} \sum_{i=1}^{M} \frac{h^{*}_{d, i} \boldsymbol{\varrho}_{i} +  \boldsymbol{\varrho}_{i} \boldsymbol{\varrho}_{i}^{\mathrm{H}} \boldsymbol{v}^{*}}
{ \left( \kappa \left| h_{d, i}+\boldsymbol{\varrho}_{i}^{\mathrm{T}} \boldsymbol{v}\right|^{2}+\frac{\sigma_{u}^{2}}{p_{b}}\right)^{2}} \right\rbrace .
\end{aligned}
\end{equation}

With the derived gradient, we can quickly obtain the sub-optimal solution of the objective function by searching in the gradient direction\footnote{
	By using the GDM, we can only obtain the numerical local optimal solution. Although the numerical solution is sufficient to offer an outstanding performance which can be seen in the simulation results later, we are still interested in the closed-form solution of the problem. To ensure the objective function is holomorphic (complex analytic), the conjugate coordinates $ \left( \boldsymbol{v}, \boldsymbol{v}^{*}\right) $ should be introduced\textcolor{blue}{\cite{zhang2017matrix}}. By using the first-order Taylor series approximation of the objective function $ f \left( \boldsymbol{v}, \boldsymbol{v}^{*}\right) $ at a given point $ \left( \boldsymbol{c}, \boldsymbol{c}^{*}\right) $, we can obtain the necessary condition for that $ \boldsymbol{c} $ is a local extreme point: the gradient vector equals to zero; By using the second-order Taylor series approximation of the objective function $ f \left( \boldsymbol{v}, \boldsymbol{v}^{*}\right) $ at a given point $  \left( \boldsymbol{c}, \boldsymbol{c}^{*}\right) $, we can obtain the necessary and sufficient condition for that $ \boldsymbol{c} $ is a local maximum point: the conjugate gradient vector equals to zero, and the full Hessian matrix, which consists of four part complex Hessian matrices, is negative semi-definite.
}. The detailed steps of the proposed GDM-based reflect beamforming algorithm is illustrated in Algorithm 1. It should be noted that the step size $ t $ is chosen via exact or backtracking line search, {i.e.}, $ t = \operatorname{argmax}_{s\geqslant0} f(\boldsymbol{\theta}+s\nabla f(\boldsymbol{\theta})) $.

\begin{algorithm}[h]
	\begin{spacing}{1.4}
		\caption{The Proposed GDM-Based Reflect Beamforming Algorithm}
		
		\textbf{Objective Function:}
		\vspace{-10 pt}
		\begin{equation*}
		f(\boldsymbol{\theta}) = 
		\sum_{i=1}^{M} \frac{\left|{h}_{d, i}+\boldsymbol{\varrho}^{\mathrm{T}}_{i} \boldsymbol{v}\right|^{2}}{\left(1+\kappa_{u}\right)\kappa_{b}\left|{h}_{d, i} + \boldsymbol{\varrho}^{\mathrm{T}}_{i} \boldsymbol{v}\right|^{2}+\frac{\sigma_{u}^{2}}{p_{b}}}.
		\end{equation*}
		\vspace{-15 pt}
		
		\textbf{Input:} The channel matrices $ \mathbf{h}_{\mathrm{d}} $, $ \mathbf{H}_{\mathrm{R}} $, and the hardware impairment coefficients $ \kappa_{b},\kappa_{u} $.
		
		Set the number of iterations $k = 0$.
		
		Set the loop stop criterion: tolerance $\epsilon > 0$.
		
		Initialize the RIS phase shifts by compensating the channels phase shifts, $\boldsymbol{\theta}^{(0)}$.
		
		Compute the objective function value under the current RIS phase shifts, $f(\boldsymbol{\theta}^{(0)})$.
		
		\While{stop criterion $\Delta f(\boldsymbol{\theta}^{(i)}) < \epsilon$ is not satisfied}{
			\text{ }Update the number of iterations $k = k +1$. 
			
			\text{ }Choose the gradient $\nabla_{\boldsymbol{\theta}} f(\boldsymbol{\theta}^{(i)})$ as the search direction. 
			
			\text{ }Choose the step size $t$ via exact or backtracking line search.
			
			\text{ }Update the optimization variable $\boldsymbol{\theta}^{(i+1)} := \boldsymbol{\theta}^{(i)} + t \nabla_{\boldsymbol{\theta}} f(\boldsymbol{\theta}^{(i)})$.
			
			\text{ }Compute the objective function difference $\Delta f(\boldsymbol{\theta}^{(i+1)}) = f(\boldsymbol{\theta}^{(i+1)}) - f(\boldsymbol{\theta}^{(i)})$.
		}
		
		\KwResult{Iteration number $k$, optimized reflect beamforming vector $\boldsymbol{\theta}$.}
	\end{spacing}
\end{algorithm}

\subsection{Asymptotic Channel Capacity}
In this subsection, we analyze the asymptotic channel capacity in the RIS-assisted communication system with hardware impairments. We consider two types of asymptotic channel capacities: the capacity as the transmit power approaches infinity ($ p_{b}, p_{u} \rightarrow \infty $), and the capacity as the numbers of antennas and reflecting elements approach infinity ($ M, N \rightarrow \infty $). In what follows, we only provide the results for the downlink channel due to space limitation. 

\begin{theorem}
	The asymptotic channel capacity $ {C}^{p_{b}}_{d}(\infty) = \lim _{p_{b} \rightarrow \infty} {C}_{d}$ is bounded as
	
	\begin{equation}
	\label{eq.12}
	\begin{aligned}
	\frac{\tau_{d}}{\tau} \log _{2}\left(1+\frac{1}{ \kappa_{b}+ \kappa_{u}(1+\kappa_{b})}\right) \leq {C}^{p_{b}}_{d}(\infty) \leq \frac{\tau_{d}}{\tau} \log _{2}\left(1+\frac{M}{ \kappa_{b}+ \kappa_{u}(M+\kappa_{b})}\right) .
	\end{aligned}
	\end{equation}	
\end{theorem}

\begin{IEEEproof}
	Assume that the phase-shifting vector of RIS is optimal. According to the Eq. (2.2) in\textcolor{blue}{\cite{silverstein1995empirical}}, the upper bound of downlink channel capacity in Eq. (\ref{downlink channel capacity}) can be rewritten as
	\begin{equation}
	\label{downlink channel capacity 2}
	{{C}}_{\mathrm{down}} 
	\leq \frac{\tau_{d}}{\tau} \mathbb{E}\left\{ \log _{2}\left(1 + \frac{\hat{\mathbf{h}}^{\mathrm{H}} \tilde{\mathbf{D}}^{-1}\hat{\mathbf{h}}}{1+\kappa_{u}\hat{\mathbf{h}}^{\mathrm{H}} \tilde{\mathbf{D}}^{-1} \hat{\mathbf{h}}}\right)\right\},
	\end{equation}		
	where $ \tilde{\mathbf{D}}=\left(1+\kappa_{u}\right) \kappa_{b} \operatorname{diag}(\hat{\mathbf{h}}\hat{\mathbf{h}}^{\mathrm{H}})+ \frac{\sigma_{u}^{2}}{p_{b}} \mathbf{I}$. Then, by applying Jensen's inequality to Eq. (\ref{downlink channel capacity 2}), we obtain the new upper bound of downlink channel capacity as
	
	\begin{equation}
	\label{downlink channel capacity 3}
	{{C}}_{\mathrm{down}} \leq \frac{\tau_{d}}{\tau}  \log _{2}\left(1 + \frac{\mathbb{E}\left\{\hat{\mathbf{h}}^{\mathrm{H}} \tilde{\mathbf{D}}^{-1}\hat{\mathbf{h}}\right\}}{1+\kappa_{u}\mathbb{E}\left\{\hat{\mathbf{h}}^{\mathrm{H}} \tilde{\mathbf{D}}^{-1} \hat{\mathbf{h}}\right\}}\right) .
	\end{equation}	
	Based on the form transformation from problem (P3) in Eq. (\ref{p3}) to problem (P4) in Eq. (\ref{p4}), the term $ \hat{\mathbf{h}}^{\mathrm{H}} \tilde{\mathbf{D}}^{-1} \hat{\mathbf{h}} $ in Eq. (\ref{downlink channel capacity 3}) can be rewritten as 
	\begin{equation}
	\label{form}
	\hat{\mathbf{h}}^{\mathrm{H}} \tilde{\mathbf{D}}^{-1} \hat{\mathbf{h}} 
	= \sum_{i=1}^{M} \frac{\left|{h}_{d, i} + \boldsymbol{\varrho}_{i}^{\mathrm{T}} \boldsymbol{v}\right|^{2}}{\left(1+\kappa_{u}\right)\kappa_{b}\left|{h}_{d, i} + \boldsymbol{\varrho}_{i}^{\mathrm{T}} \boldsymbol{v}\right|^{2}+\frac{\sigma_{u}^{2}}{p_{b}}} .
	\end{equation}
	As the transmit power $ p_{b} $ at the BS approaches infinity, the term  $ \frac{\sigma_{u}^{2}}{p_{b}} $ in the denominator of Eq. (\ref{form}) approaches zero, and thus we have 
	\begin{equation}
	\label{eq.14}
	\lim _{p_{b} \rightarrow \infty} \mathbb{E}\left\{\hat{\mathbf{h}}^{\mathrm{H}} \tilde{\mathbf{D}}^{-1} \hat{\mathbf{h}}\right\} = \frac{M}{\left(1+\kappa_{u}\right)\kappa_{b}}.
	\end{equation}
	Then, the upper bound in Eq. (\ref{eq.12}) is achieved by substituting Eq. (\ref{eq.14}) into Eq. (\ref{downlink channel capacity 3}). The lower bound is asymptotically obtained by using $ \mathbb{E}\left\{\boldsymbol{x} \boldsymbol{x}^{\mathrm{H}}\right\} = \frac{p_{b}}{M} \mathbf{I}$.
\end{IEEEproof}

\begin{theorem}
	The asymptotic channel capacity $C^{M,N}_{d}(\infty) = \lim _{M,N \rightarrow \infty} C_{d} $ is bounded as 
	\begin{equation}
	\label{theroem 4}
	\begin{aligned}
	\frac{\tau_{d}}{\tau} \log _{2}\left(1+\frac{1}{ \kappa_{b}+ \kappa_{u}(1+\kappa_{b})}\right) \leq C^{M,N}_{d}(\infty) &\leq \frac{\tau_{d}}{\tau} \log _{2}\left(1+\frac{1}{\kappa_{u}}\right) .
	\end{aligned}
	\end{equation}
\end{theorem}
\begin{IEEEproof}
	Based on the form of $\hat{\mathbf{h}}^{\mathrm{H}} \tilde{\mathbf{D}}^{-1}\hat{\mathbf{h}}$ in Eq. (\ref{form}), we observe that it approaches infinity as the numbers of antennas $M$ and reflecting elements $N $ approach infinity, i.e.,
	\begin{equation}
	\label{form2}
	\lim _{M, N \rightarrow \infty} \mathbb{E}\left\{\hat{\mathbf{h}}^{\mathrm{H}} \tilde{\mathbf{D}}^{-1} \hat{\mathbf{h}}\right\} \rightarrow \infty .
	\end{equation}
	Then, by substituting Eq. (\ref{form2}) into Eq. (\ref{downlink channel capacity 3}), the upper bound of channel capacity is obtained in Eq. (\ref{theroem 4}). The lower bound is asymptotically achieved by using $\mathbb{E}\left\{\boldsymbol{x} \boldsymbol{x}^{\mathrm{H}}\right\} = \frac{p_{b}}{M} \mathbf{I}.$
\end{IEEEproof}

\begin{remark}
	When the BS antenna number is finite and the reflecting element number tends to infinity, the upper bound of asymptotic channel capacity is the same as that of the case where the transmit power tends to infinity. This implies that the major function of RIS is to reach the channel capacity limit at low transmit power. That is also the reason why RIS can improve the energy and spectrum efficiency of wireless communication systems without the need of complex processing on signals. However, we cannot increase the asymptotic channel capacity assisted by the RIS when the BS antenna number is fixed.    
\end{remark}

\subsection{Numerical Results}
In this subsection, we evaluate the performance of our proposed beamforming optimization algorithm, and characterize the asymptotic channel capacities we derived in Theorems 3 and 4.

We assume that the BS is equipped with $M=5$ antennas, the transmit power is $15$ dB, the impairment coefficients are set as $\kappa_{b} = \kappa_{u} = 0.01^2$, and the transmit beamforming is optimized by the solution we derived in Theorem 2.
Then, by utilizing the Monte Carlo trials, we compare the downlink spectral efficiency in the case of optimized phase shifts and that in the case of no phase shift versus the reflecting element number $N$ and SNR which is defined as $\frac{p_{b}}{\sigma_{u}^2}$ in \textcolor{blue}{Fig. 6 and 7}, respectively. It shows that when the reflecting element number $N$ in the range of $[5, 100]$, the reflect beamforming gain is significant. 
However, if the reflecting element number $N>100$, the reflect beamforming gain degrades with the increase of $N$. It is because the existence of hardware impairments, the downlink spectral efficiency increases with the number of reflecting elements and gradually reaches the limit, and the reflect beamforming gain gradually degrades. Similarly, the reflect beamforming gain is significant in an appropriate range of SNR. Thus, if we utilize an appropriate number of reflecting elements, the reflect beamforming can bring a significant improvement in spectral efficiency, while if we utilize massive reflecting elements or the transmit power is high, it is not very necessary to optimize the reflect beamforming.

\begin{figure}[htbp]
	\centering
	\begin{tabular}{cc}
		\begin{minipage}[t]{0.472\linewidth}
			\flushleft
			\includegraphics[width = 1\linewidth]{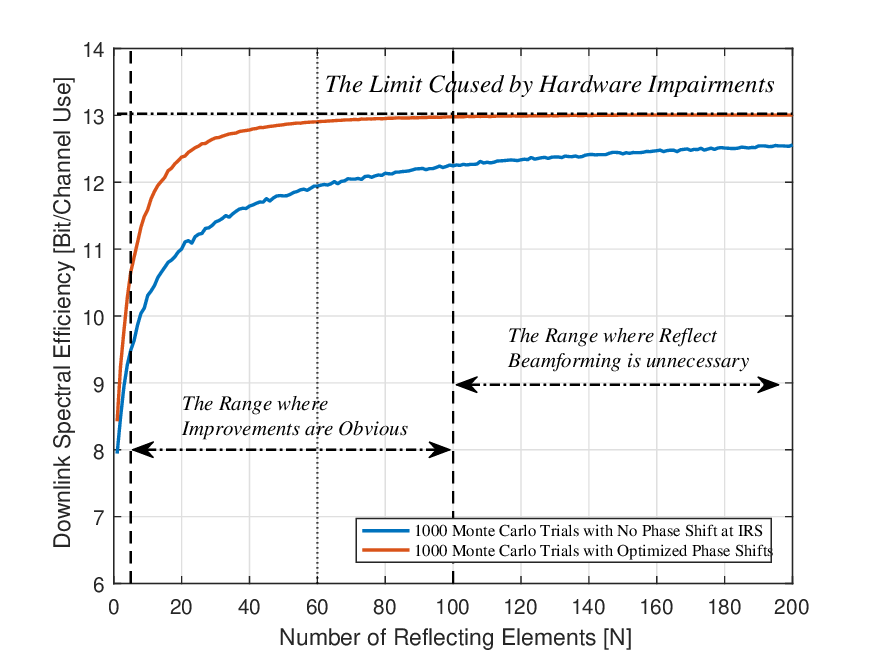}
			\vspace{-23 pt}
			\caption{The optimized downlink spectral efficiency versus the number of reflecting elements.}
		\end{minipage}
		\;\;
		\begin{minipage}[t]{0.472\linewidth}
			\includegraphics[width = 1\linewidth]{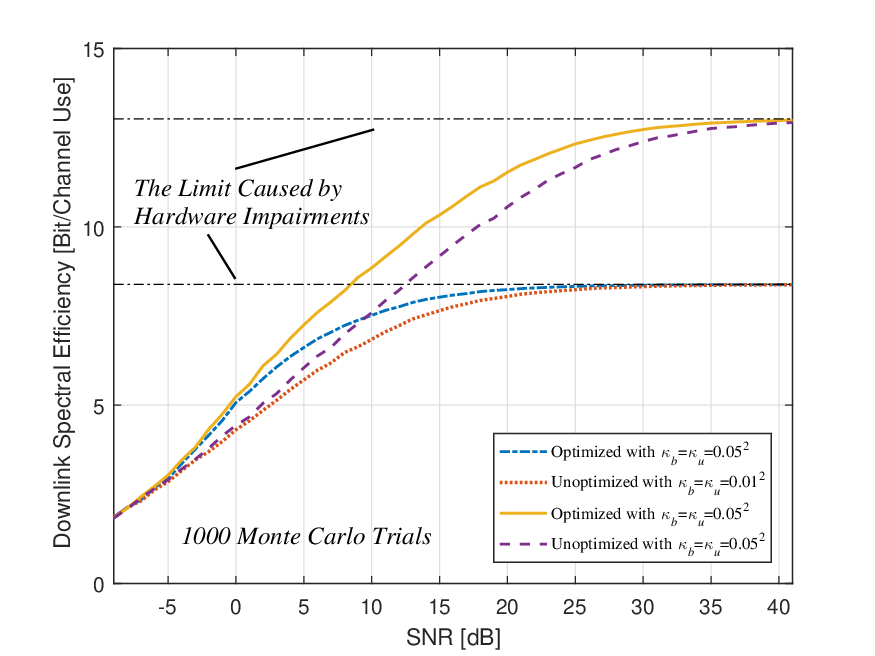}
			\vspace{-23 pt}
			\caption{The optimized downlink spectral efficiency versus the SNR with different impairment coefficients.}
		\end{minipage}
	\end{tabular}
\end{figure}

\begin{figure}[htbp]
	\vspace{-10 pt}
	\centering
	\begin{tabular}{cc}
		\begin{minipage}[t]{0.472\linewidth}
			\flushleft
			\includegraphics[width = 1\linewidth]{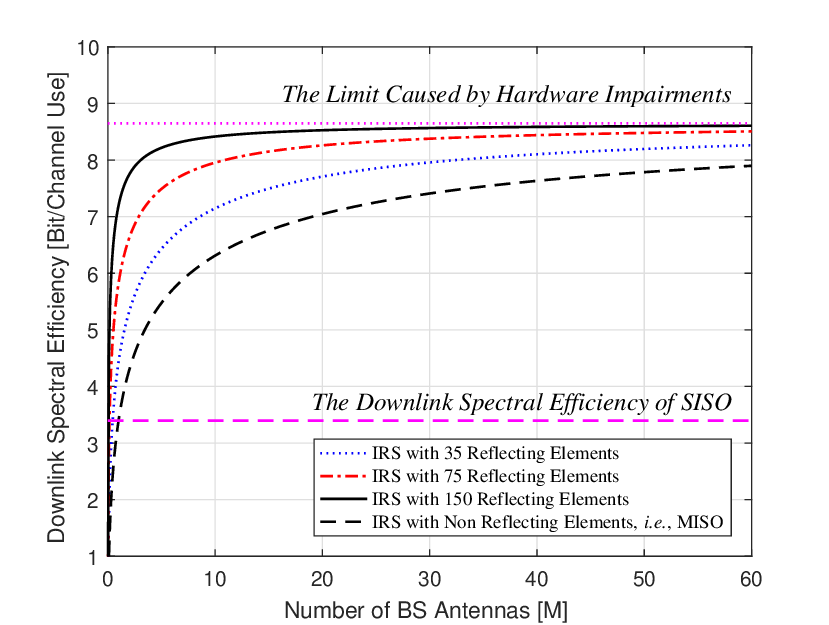}
			\vspace{-23 pt}
			\caption{The downlink spectral efficiency versus the number of BS antennas with different numbers of reflecting elements.}
		\end{minipage}
		\;\;
		\begin{minipage}[t]{0.472\linewidth}
			\includegraphics[width = 1\linewidth]{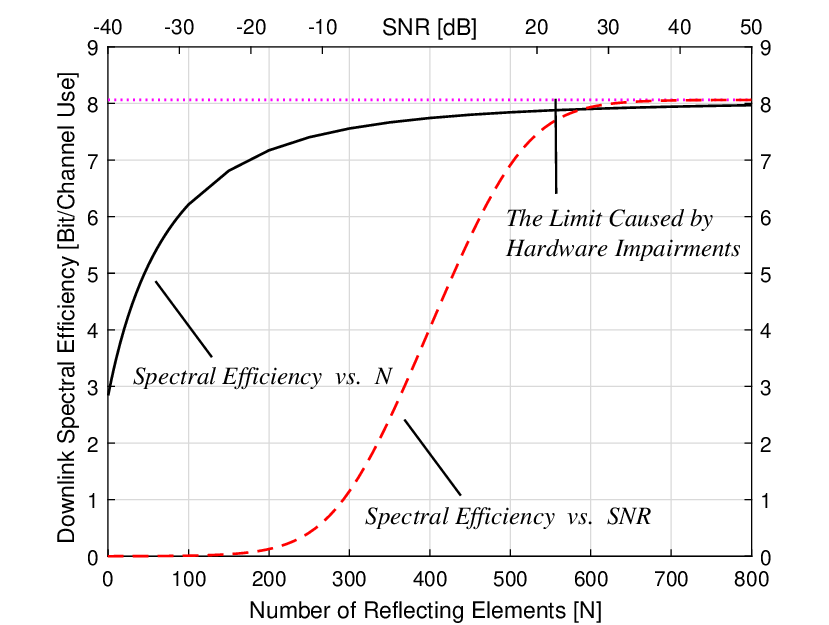}
			\vspace{-23 pt}
			\caption{The downlink spectral efficiency versus the SNR, and versus the number of reflecting elements.}
		\end{minipage}
	\end{tabular}
\vspace{-10 pt}
\end{figure}

Then, we characterize the asymptotic channel capacities. \textcolor{blue}{Fig. 8} shows the downlink spectral efficiency as a function of the BS antenna number $M$ with different reflecting element numbers $N$. We observe that the downlink spectral efficiency increases with $M$ and $N$, and converges to a finite limit caused by hardware impairments. In particular, when the number of BS antennas $M$ reaches a certain value, the growth of spectral efficiency that continues to increase $M$ becomes less noticeable. This observation is encouraging for that it is not very necessary to cost a lot on expensive antennas, which corresponds to the requirements of new communication paradigms. In \textcolor{blue}{Fig. 9}, we fix the number of BS antennas, and plot the downlink spectral efficiency versus the reflecting element number $ N $ and SNR that is defined as $ \frac{p_{\mathrm{b}}}{\sigma^{2}_{\mathrm{u}}}$, respectively. From \textcolor{blue}{Fig. 9}, it is observed that the downlink spectral efficiency increases with $N$ and SNR, and two curves have the same converged value which caused by hardware impairments. This observation confirms the fact mentioned in Remark 2 that the major function of RIS is to reach the channel capacity limit using lower transmit power.

\section{Power Scaling Law and Energy Efficiency}
\subsection{Power Scaling Law at User}
Prior researches\textcolor{blue}{\cite{6415388, 6457363, 6902790}} show that the transmit power can be reduced without degrading the communication performance by using multiple antennas, e.g., a transmitter with $M$ antennas can reduce the transmit power as $\frac{1}{M^{\alpha}}$ where $0<\alpha<\frac{1}{2}$, and still achieve non-zero spectral efficiency as $M$ tends to infinity\textcolor{blue}{\cite{6457363}}. Encouraged by this advantage, in this subsection, we quantify the power scaling law for RIS-assisted wireless systems with hardware impairments in both cases of perfect CSI and estimated CSI. Theoretically, the maximum-ratio combining detector achieves fairly well performance\textcolor{blue}{\cite{5595728, 6457363}}, thus we restrict consideration to this detector. The channels of BS-user link $\mathbf{h}_{\mathrm{d}}$, BS-RIS link $\mathbf{G}$ and RIS-user link $\mathbf{h}_{\mathrm{r}}$ are mutually independent matrices whose elements are independent and identically distributed (i.i.d.) zero-mean random variables. Then, according to the law of large numbers, we have 
\begin{equation}
\label{eq.15}
\frac{1}{M} \mathbf{h}_{\mathrm{d}}^{\mathrm{H}} \mathbf{h}_{\mathrm{d}} \rightarrow \beta_{d}^{2} \; \text{,} \; \text { as } M \rightarrow \infty , 
\end{equation}
\begin{equation}
\label{lawoflargenumber}
\frac{1}{MN^{2}}\left(\mathbf{G} \mathbf{h}_{\mathrm{r}}\right)^{\mathrm{H}}\left(\mathbf{G} \mathbf{h}_{\mathrm{r}}\right) \rightarrow \beta_{r}^{2} \; \text{,} \; \text{ as } N \rightarrow \infty ,
\end{equation}
where $ \beta_{d}^{2} = \mathbb{E}\lbrace |{h}_{d,i}|^{2}\rbrace $, $ \beta_{r}^{2} = \mathbb{E}\lbrace|{G}_{i, j}|^{2}\rbrace \mathbb{E}\lbrace|{h}_{r, j}|^{2}\rbrace $, and $ {h}_{d,i} $, $ {g}_{i,j} $ and $ {h}_{r,j} $ are the entries of $ \mathbf{h}_{\mathrm{d}} $, $\mathbf{G}$ and $ \mathbf{h}_{\mathrm{r}} $, respectively.

\vspace{5 pt}
\subsubsection{BS with perfect CSI}
With perfect CSI, the detector vector at the BS is $\mathbf{A} = \mathbf{h}_{\mathrm{d}}+\mathbf{G}\mathbf{\Phi}\mathbf{h}_{\mathrm{r}}$ when using MRC. As illustrated in Section III, the phase noise at the RIS is random and unknown to the BS, thus the detector vector is $ \mathbf{h}_{\mathrm{d}}+\mathbf{G}\mathbf{\Phi}\mathbf{h}_{\mathrm{r}} $ rather than $ \mathbf{h}_{\mathrm{d}}+\mathbf{G}\hat{\mathbf{\Phi}}\mathbf{h}_{\mathrm{r}} $. The transmitted data signal can be detected by multiplying the received signal $\boldsymbol{y} $ with $ \mathbf{A}^{\mathrm{H}} $, {i.e.}, the detected signal is $r=\mathbf{A}^{\mathrm{H}}\boldsymbol{y}$ which can be rewritten as follows:
\begin{equation}
r=\mathbf{h}^{\mathrm{H}}\hat{\mathbf{h}}\left(x+\eta_{u}\right)+\mathbf{h}^{\mathrm{H}}\boldsymbol{\eta}_{b}+\mathbf{h}^{\mathrm{H}}\boldsymbol{n} ,
\end{equation}
where $\mathbf{h}$ represents $\mathbf{h}_{\mathrm{d}}+\mathbf{G}\mathbf{\Phi}\mathbf{h}_{\mathrm{r}}$ and $\hat{\mathbf{h}}$ represents $\mathbf{h}_{\mathrm{d}}+\mathbf{G}\hat{\mathbf{\Phi}}\mathbf{h}_{\mathrm{r}}$, for simplicity. It should be noted that the expectation of $ \Delta \theta_{i} $ is zero. Then, the achievable rate $ \mathcal{R}_{u}^{p} $ of the uplink channel can be obtained as follows:
\begin{equation}
\label{R_perfect}
\mathcal{R}_{u}^{p}=\mathbb{E}\left\{\log _{2}\left(1+\frac{p_{u}\left|\mathbf{h}^{\mathrm{H}}\mathbf{h}\right|^{2}}{\kappa_{u}p_{u}\left|\mathbf{h}^{\mathrm{H}}\mathbf{h}\right|^{2}+\left\|\mathbf{h}\right\|_{2}^{2}\left(\sigma_{b}^{2}+p_{u}\kappa_{b}\left(1+\kappa_{u}\right)\right)}\right)\right\}   .
\end{equation}

\begin{theorem}
	Assuming that the BS has perfect channel state information and the transmit power of the user is scaled with the BS antenna number $ M $ and the RIS reflecting element number $ N $ according to $p_{u}={{E}_{u}}/{(M+k M N^{2})}$ where $ {E}_{u} $ is fixed and $ k= {\beta_{r}^{2}}/{ \beta_{d}^{2}}$, we have
	\begin{equation}
	\label{eq.21}
	\mathcal{R}_{u}^{p} \rightarrow \log _{2}\left(1+\frac{{E}_{u}\beta_{d}^{2}}{\kappa_{u} {E}_{u} \beta_{d}^{2}+\sigma_{b}^{2}}\right) \text{,} \; \text{ as } M, N \rightarrow \infty .
	\end{equation}
\end{theorem}

\begin{IEEEproof}
	By substituting $p_{u}={{E}_{u}}/{( M+k M N^{2})}$ into Eq. (\ref{R_perfect}) and using the law of large numbers reviewed in Eqs. (\ref{eq.15}) and (\ref{lawoflargenumber}), we obtain the asymptotic achievable rate limit $ \mathcal{R}_{u}^{p} $ as $M, N \rightarrow \infty$ in Eq. (\ref{eq.21}).
\end{IEEEproof}

\vspace{5 pt}
\subsubsection{BS with imperfect CSI}
In practice, the BS has to estimate the channels, and there exists estimation error as we discussed in Section III. For simplicity, we denote the channel estimation error as $\boldsymbol{\mathcal{E}}={\hat{\mathbf{h}}}^{\prime} - \hat{\mathbf{h}}$. Referring to Eq. (33) in\textcolor{blue}{\cite{6457363}}, the elements of $\boldsymbol{\mathcal{E}}$ are random variables with zero means and variances $\frac{\beta}{p_{u} \beta+1}$ where $\beta=(1+k N^{2})\beta_{d}^{2}$. We rewrite the received signal as 
\begin{equation}
r= {\hat{\mathbf{h}}}^{\prime \mathrm{H}} \left( {\hat{\mathbf{h}}}^{\prime} - \boldsymbol{\mathcal{E}}\right) \left(x+\eta_{u}\right)+{\hat{\mathbf{h}}}^{\prime \mathrm{H}} \boldsymbol{\eta}_{b} + {\hat{\mathbf{h}}}^{\prime \mathrm{H}} \boldsymbol{n} .
\end{equation}
Similarly to the Eq. (38) in\textcolor{blue}{\cite{6457363}}, the achievable rate $ \mathcal{R}_{u}^{ip} $ of the uplink channel is given as follows where each element of ${\hat{\mathbf{h}}}^{\prime}$ is random with zero mean and variance $\frac{p_{u} \beta^{2}}{p_{u} \beta+1}$,
\begin{equation}
\label{R_imperfect}
\mathcal{R}_{u}^{ip}=\mathbb{E}\left\{\log _{2}\left(1+\frac{p_{u}\left|{\hat{\mathbf{h}}}^{\prime \mathrm{H}} {\hat{\mathbf{h}}}^{\prime} \right|^{2}}{\left(1+\kappa_{u}\right) p_{u}\left\|{\hat{\mathbf{h}}}^{\prime}\right\|_{2}^{2}\frac{\beta}{p_{u} \beta+1}+\kappa_{u}p_{u}\left|{\hat{\mathbf{h}}}^{\prime \mathrm{H}} {\hat{\mathbf{h}}}^{\prime}\right|^{2}+\left\|{\hat{\mathbf{h}}}^{\prime}\right\|_{2}^{2}\left(\sigma_{b}^{2}+p_{u}\kappa_{b}\left(1+\kappa_{u}\right)\right)}\right)\right\} .
\end{equation}

\begin{theorem}
	Assuming that the BS has imperfect channel state information and the transmit power of the user is scaled with the BS antenna number $ M $ and the RIS reflecting element number $ N $ according to $p_{u}={{E}_{u}}/{(\sqrt{M}(1+k N^{2})) }$ where $ {E}_{u} $ is fixed and $ k= {\beta_{r}^{2}}/{ \beta_{d}^{2}}$, we have
	\begin{equation}
	\label{eq.23}
	\mathcal{R}_{u}^{ip} \rightarrow \log _{2}\left(1+\frac{{E}^{2}_{u}\beta_{d}^{4}}{\kappa_{u} {E}^{2}_{u}\beta_{d}^{4}+\sigma_{b}^{2}} \right)\; \text{,} \; \text{ as } M,N \rightarrow \infty.
	\end{equation}
\end{theorem}

\begin{IEEEproof}
	The proof follows the similar procedures with Proposition 2. By substituting $p_{u}={{E}_{u}}/{(\sqrt{M}(1+k N^{2}))}$ into Eq. (\ref{R_imperfect}), and using the law of large number reviewed in Eqs. (\ref{eq.15}) and (\ref{lawoflargenumber}) along with the variances of elements in the estimated channel vector $\mathbf{h}^{\prime} $ and the estimation error vector $ \boldsymbol{\mathcal{E}} $, we obtain the limit of the achievable rate $ \mathcal{R}_{u}^{ip} $ as $ M,N \rightarrow \infty$ in Eq. (\ref{eq.23}). 
\end{IEEEproof}

\begin{remark}
	Theorem 5 shows that if the BS has perfect channel state information, and $ M $ and $ N $ are large enough, the performance of an RIS-assisted communication system with $ M $-antenna BS, $ N $-reflecting element RIS and the transmit power $ {{E}_{u}}/ {(M(1+k N^{2}))} $ at the user is equal to the performance of a single-input single-output (SISO) system with the transmit power ${E}_{u}$ at the user. 
	Theorem 6 shows that if the BS has estimated channel state information, and $ M $ and $ N $ are large enough, the performance of an RIS-assisted communication system with $ M $-antenna BS, $ N $-reflecting element RIS and the transmit power $ {{E}_{u}}/ {(\sqrt{M}(1+k N^{2}))} $ at the user is equal to the performance of a SISO system with the transmit power ${E}^{2}_{u}\beta_{d}^{2}$ at the user. 
	Theorem 6 also implies that the transmit power can be cut proportionally to ${{E}_{u}}/{(M^{\alpha}(1+k N^{2})^{2\alpha})}$ where the parameter $\alpha \leq \frac{1}{2}$. If the parameter $\alpha>\frac{1}{2}$, the achievable rate of the uplink channel will converge towards zero as $M \rightarrow \infty$ and $N \rightarrow \infty$.
\end{remark}

\vspace{-15 pt}
\subsection{Energy Efficiency of Downlink}
In this subsection, we analyze the energy efficiency which is measured by Bit/Joule, and a common definition is the ratio of the spectral efficiency (in Bit/Channel Use) to the transmit power (in Joule/Channel Use). The energy consumed at the BS and the user (per coherence period) is 
\begin{equation}
{E}=\tau_{d\;} p_{b} + \left(\tau_{{pilot}}+ \tau_{u}\right) p_{u} ,
\end{equation}
where $\tau_{{pilot}}$, $\tau_{u}$ and $\tau_{d}$ are the consumed time of pilot-based channel estimation, uplink transmission and downlink transmission, respectively.
In addition, there exists a baseband circuit power consumption which can be modeled as $ M\rho + \zeta $\textcolor{blue}{\cite{6056691,6251827}}. The parameter $ \rho \geq 0 $ describes the circuit power which scales with the BS antenna number $M$. The parameter $ \zeta > 0 $ describes the circuit power which is static. Then, the average power (in Joule/Channel Use) can be given as 
\begin{equation}
\label{eq.17}
\begin{aligned}
\frac{{E}}{\tau} =  \underbrace{\left( \frac{\tau_{d}}{\tau_{u}+\tau_{d}} \left(  \frac{\tau_{{pilot\;}}p_{u}}{\tau} + M\rho + \zeta\right)  + \frac{\tau_{d\;} p_{b}}{\tau} \right) }_{\text{Downlink Average Power}}
+ \underbrace{\left( \frac{\tau_{u}}{\tau_{u}+\tau_{d}} \left(  \frac{\tau_{{pilot\;}}p_{u}}{\tau} + M\rho + \zeta \right) +  \frac{\tau_{u\;} p_{u}}{\tau} \right)}_{\text{Uplink Average Power}}.
\end{aligned}
\end{equation}
The first term of Eq. (\ref{eq.17}) refers to the average power of the downlink transmission. Based on the power consumption model established above, we give the definition of the overall energy efficiency as follows:

\begin{definition}
	The energy efficiency of downlink is 
	\begin{equation}
	\label{eq.18}
	\Xi_{\mathrm{down}} = \frac{{C}_{d}}{\frac{\tau_{d}}{\tau_{u}+\tau_{d}}\left(\frac{\tau_{{pilot\;}}p_{u}}{\tau} + M\rho + \zeta \right) + \frac{\tau_{d\;} p_{b}}{\tau}} ,
	\end{equation}
	where $ {C}_{d} $ is the channel capacity of downlink.
\end{definition}

\begin{theorem}
	Suppose we want to maximize the energy efficiency of downlink with respect to the transmit power ($p_{b} , p_{u} \geq 0$), the number of BS antennas ($M \geq 0 $), and the number of reflecting elements ($N \geq 0$). If the parameter $ \rho=0 $, the maximal energy efficiency $\Xi^{\mathrm{max}}_{\mathrm{down}}$ is bounded as 
	\begin{equation}
	\label{eq.19}
	\frac{ \log _{2}\left(1+\frac{1}{ \kappa_{b}+ \kappa_{u}(1+\kappa_{b})}\right)}{\frac{\tau\zeta}{\tau_{u}+\tau_{d}}}\leq  \Xi^{\mathrm{max}}_{\mathrm{down}} \leq \frac{\log _{2}\left(1+\frac{1}{\kappa_{u}}\right)}{\frac{\tau\zeta}{\tau_{u} + \tau_{d}}} .
	\end{equation}
	If the parameter $ \rho>0 $, the upper bound is still valid, but the asymptotic energy efficiency is zero, i.e., the maximal energy efficiency can be achieved at certain finite $ M $.
\end{theorem}

\begin{IEEEproof}
	Based on the definition of energy efficiency given in Definition 1 and the asymptotic channel capacity of downlink given in Theorem 4, we can prove Theorem 7. We maximize the energy efficiency with respect to the transmit power, the number of BS antennas, and the number of reflecting elements: 1) by neglecting the transmit power terms in the denominator of Eq. (\ref{eq.18});  2) and applying the bounds of the asymptotic channel capacity limit in Theorem 4 to the numerator of Eq. (\ref{eq.18}). Then, we obtain the upper bound and the lower bound of the maximal energy efficiency of downlink in Eq. (\ref{eq.19}).
\end{IEEEproof}

\begin{remark}
	If the number of BS antennas is fixed and the circuit power have no correlation with the number of BS antennas, i.e., the parameter $ \rho=0 $, increasing the number of reflecting elements on RIS can only make the energy efficiency close to the upper bound given as follows,
	\begin{equation}
	\label{eq.20}
	\Xi^{\mathrm{max}}_{\mathrm{down}} \leq \frac{\log _{2}\left(1+\frac{M}{ \kappa_{b}+ \kappa_{u}(M+\kappa_{b})}\right)}{\frac{\tau\zeta}{\tau_{u} + \tau_{d}}} .
	\end{equation}
	The upper bound of maximal energy efficiency only can be increased from Eq. (\ref{eq.20}) to Eq. (\ref{eq.19}) by increasing the number of BS antennas.
	
	If the circuit power scales with the number of BS antennas, i.e., the parameter $ \rho>0 $, the maximal energy efficiency only can be achieved at a certain finite $ M $, which depends on the parameters $ \rho $ and $ \zeta $. Meanwhile, the effect of increasing the reflecting elements in this case is still to make the energy efficiency close to the upper bound under the current $ M $.
\end{remark}

\vspace{-15 pt}
\subsection{Numerical Results}
In this subsection, we numerically illustrate the power scaling law we derived in Theorems 5 and 6, and the maximal energy efficiency we derived in Theorem 7. 

\textcolor{blue}{Fig. 10} compares the uplink spectral efficiency of RIS-assisted system with that of MISO and SISO system versus SNR in both cases of perfect and imperfect CSI. In this figure, the BS is equipped with $ M=20 $ antennas, the RIS is equipped with $ N=100$ reflecting elements, and $ \kappa_{b}=\kappa_{u}=0.05^2$. It shows that the user in RIS-assisted system can obtain the same spectral efficiency at lower transmit power than that in MISO and SISO system, which verifies the correctness of our results in Theorems 5 and 6. 

Then, to illustrate the maximal energy efficiency, we consider the case where the reflecting element number $N$ approaches infinity. In this case, the upper bound of asymptotic channel capacity is same with that of the case where the transmit power $p_{b}$ approaches infinity, as been discussed in Section IV. Thus, we neglect the transmit power terms in the energy efficiency. It is reasonable because the energy efficiency can be improved by reducing the transmit power without the reduction of spectral efficiency.
To illustrate the difference between the static circuit power and the circuit power which scales with the number of BS antennas $M$, we consider four different splittings between $ \rho $ and $ \zeta $: $ \frac{\rho}{\rho+\zeta} \in \{0, 0.002, 0.01, 0.02\} $. 
Based on the power consumption numbers reported in\textcolor{blue}{\cite{auer2010d2}}, we consider the power consumption that $ \rho+\zeta = 0.5\mu \; \text{Joule/Channel Use} $. 
\textcolor{blue}{Fig.~11} shows the maximal energy efficiency versus the BS antenna number $M$ with different splitting between $ \rho $ and $ \zeta$. 
It is observed that when $ \rho = 0 $, {i.e.}, the circuit power is static, the maximal energy efficiency increases with the BS antenna number $M$ and converges to a finite value, which conforms to the upper bound derived in Theorem 7. While $ \rho > 0 $, {i.e.}, some part of power consumption scales with $ M $, the maximal energy efficiency can be achieved at some finite $ M $. When the part of power consumption which scales with $ M $ is non-negligible, {e.g.}, $\frac{\rho}{\rho+\zeta} = 0.01, 0.02$ in \textcolor{blue}{Fig.~11}, the BS antenna number is unnecessary to be very large. This conclusion is similar to the result in \textcolor{blue}{Fig.~8}, which implies that an RIS-assisted system can achieve both high spectral efficiency and high energy efficiency with a small number of BS antennas. As to the number of reflecting elements at RIS, it should be as large as possible without degrading the performance, {e.g.}, channel estimation accuracy, bit error rate. This conclusion is encouraging as it implies that there is no need to cost a lot on expensive high-quality antennas when the system is assisted by an RIS. This corresponds to the requirements of new communication paradigms.
\vspace{-7 pt}
\begin{figure}[htbp]
	\centering
	\begin{tabular}{cc}
		\begin{minipage}[t]{0.472\linewidth}
			\flushleft
			\includegraphics[width = 1\linewidth]{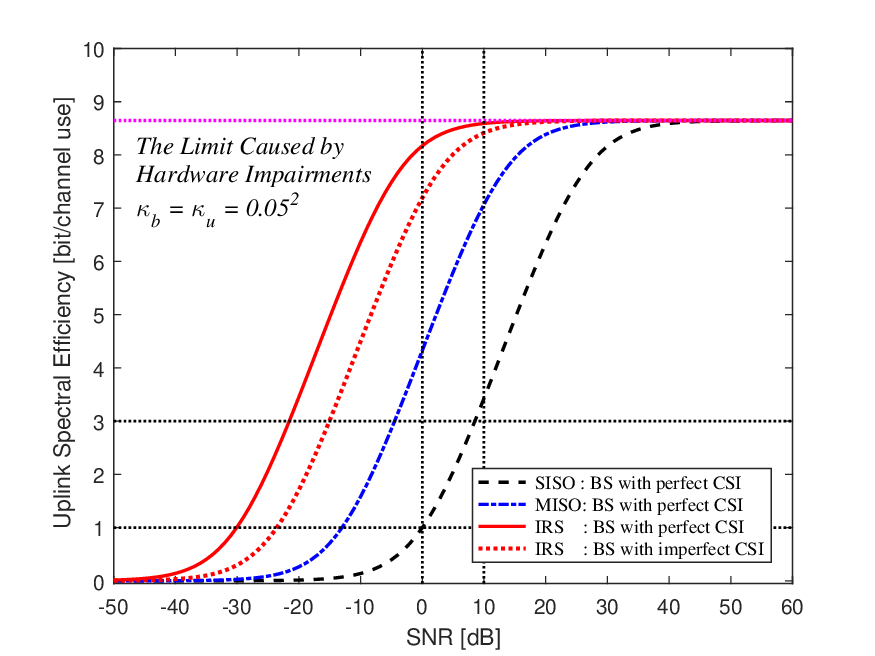}
			\vspace{-23 pt}
			\caption{The uplink spectral efficiency versus the SNR in SISO, MISO, and RIS systems, respectively.}
		\end{minipage}
		\;\;
		\begin{minipage}[t]{0.472\linewidth}
			\includegraphics[width = 1\linewidth]{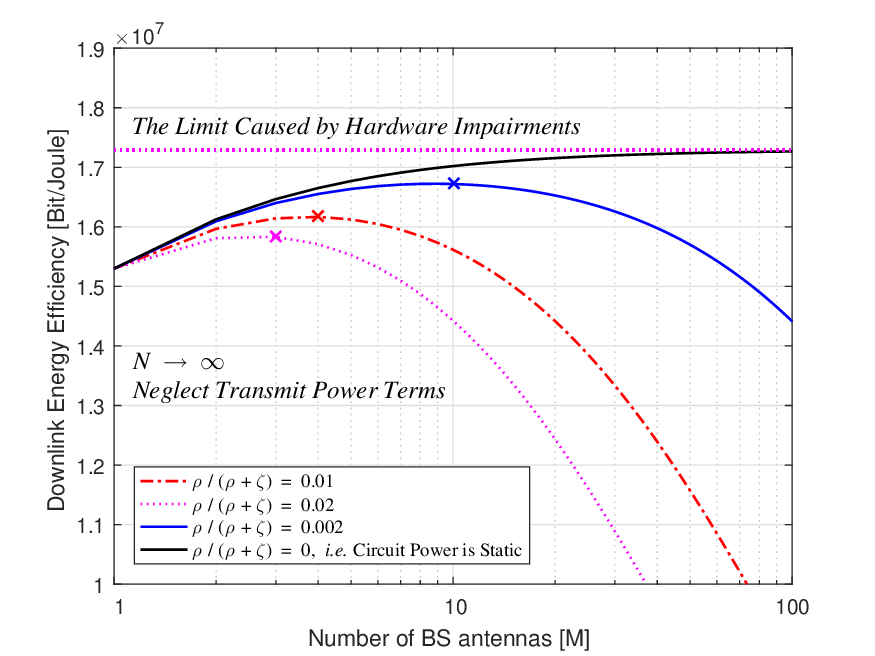}
			\vspace{-23 pt}
			\caption{The maximal energy efficiency of downlink versus the number of BS antennas with different $ \rho $ and $ \zeta $.}
		\end{minipage}
	\end{tabular}
\end{figure}

\section{Conclusion}
In this paper, we consider an RIS-assisted wireless communication system with non-ideal hardware. We characterize and evaluate  the system performance in this case comprehensively, which includes channel estimation accuracy, spectrum and energy efficiency, channel capacity and power scaling law, and analyze the impacts of hardware impairments on them. We also give the solutions to reduce these impacts and to optimize the joint transmit and reflect beamforming. The validity of all the theorems and corollaries derived in this paper are verified by simulations. Our experimental results also reveal an encouraging phenomenon that the RIS-assisted wireless system with massive reflecting elements can achieve both high spectrum and energy efficiency without the needs of massive antennas and allocating too many resources to optimize the reflect beamforming, especially when the hardware impairments are non-negligible.

\end{spacing}

\begin{spacing}{1.425}

\bibliographystyle{IEEEtran} 
\bibliography{reference}

\end{spacing}

\end{document}